# Growth of Highly Conductive $PtSe_2$ Films Controlled by Raman Metrics for High-Frequency Photodetectors and Optoelectronic Mixers at 1.55 µm

*Eva Desgué*[1*‡], *Ivan Verschueren*[1†], *Marin Tharrault*[2†], *Djordje Dosenovic*[3], *Ludovic Largeau*[4], *Eva Grimaldi*[1†], *Delphine Pommier*[1], *Doriane Jussey*[1], *Bérangère Moreau*[3], *Dominique Carisetti*[1], *Laurent Gangloff*[1], *Patrick Plouhinec*[1], *Naomie Messudom*[1†], *Zineb Bouyid*[1†], *Didier Pribat*[5], *Julien Chaste*[4], *Abdelkarim Ouerghi*[4], *Bernard Plaçais*[2], *Emmanuel Baudin*[2], *Hanako Okuno*[3‡], *Pierre Legagneux*[1*‡]

[1] Thales Research & Technology (TRT), 1 avenue Augustin Fresnel, 91120 Palaiseau, France

[2] Laboratoire de Physique de l'École Normale Supérieure (LPENS), 24 Rue Lhomond, 75005 Paris, France

[3] CEA Grenoble, 17 avenue des Martyrs, 38000 Grenoble, France

[4] Center for Nanoscience and Nanotechnology (C2N), 10 boulevard Thomas Gobert, 91120 Palaiseau, France

[5] Laboratoire de Physique des Interfaces et Couches Minces (LPICM), Ecole Polytechnique, 91120 Palaiseau, France

* Email: eva.desgue@thalesgroup.com , pierre.legagneux@thalesgroup.com

‡ *E.D., H.O. and P.L. contributed equally to this work*

**Abstract**: Two-dimensional $PtSe_2$ exhibits outstanding intrinsic properties such as high carrier mobility, tunable bandgap, broadband absorption and air stability, making it ideal for (opto)electronic applications. In particular, thick $PtSe_2$ is semimetallic and well suited for ultrafast optoelectronics in the infrared domain. However, achieving $PtSe_2$ films of high crystalline quality with controlled properties on low-cost and insulating substrates remains challenging. Here, highly crystalline semimetallic $PtSe_2$ films are grown by molecular beam epitaxy on sapphire substrates. It is shown how an optimized post-growth annealing remarkably improves the out-of-plane crystallinity and leads to record sheet conductances, up to 1.6 mS. In-depth structural analyses reveal the strong influence of the domain arrangement within the films on their electrical properties. Films that are mainly composed of vertically single crystalline domains exhibit high sheet conductance (1.1 - 1.6 mS), whereas films that contain superimposed twisted domains present low sheet conductance (0.5 - 0.6 mS). Moreover, it is demonstrated that the $A_{1g}$ Raman peak width, in addition to the commonly used $E_g$ peak width, are both effective metrics for evaluating the quality of $PtSe_2$: films with narrower $E_g$ and $A_{1g}$ peaks exhibit higher in-plane and out-of-plane crystalline quality, respectively, as well as higher sheet conductance. Finally, coplanar waveguides integrating a semimetallic $PtSe_2$ channel are fabricated on a 2-inch sapphire substrate to demonstrate optoelectronic devices operating at the 1.55 µm telecom wavelength. This includes photodetectors with a record 60 GHz bandwidth and the first $PtSe_2$-based optoelectronic mixer with a bandwidth above 30 GHz.

Transition metal dichalcogenides (TMDs) are highly promising two-dimensional (2D) materials for (opto)electronic applications.[1,2,3,4] They exhibit thickness-dependent electronic properties, broadband absorptions, strong light-matter interaction per 2D layer and can be integrated onto different platforms,[5,6] including CMOS and photonic platforms.[7,8,9,10] Platinum diselenide ($PtSe_2$) is an original TMD which is semiconducting in a few layers and semimetallic in thicker films. It also stands out because it can absorb in the infrared (IR), presents one of the highest theoretically predicted carrier mobilities (> 4000 $cm^2.(V.s)^{-1}$),[11] with experimentally reported values above 200 $cm^2.(V.s)^{-1}$ at room temperature, and is highly air stable.[12,13,14,15,16] Therefore, $PtSe_2$ is particularly suitable for IR optoelectronics.[17,18,19,20]

$PtSe_2$-based photodetectors operating at the telecom wavelength of 1.55 µm have been widely studied.[19,21,22,23] However, only a few studies have demonstrated the high-frequency (> 1 GHz) bandwidth required for high-speed telecommunications: based on transferred $PtSe_2$ films grown by vapor phase transport (VPT), Wang *et al.* demonstrated 17 GHz bandwidth photodetectors fabricated on a $SiO_2$/Si substrate[24] and 35 GHz bandwidth waveguide photodetectors realized on a silicon nitride platform.[25] In these experiments, semimetallic $PtSe_2$ acts as a photoconductor as the light absorption leads to an increase in carrier density and, subsequently, an increase in conductance.[12] While devices based on semimetallic photoconductors are not competitive in terms of detectivity, they are particularly well suited for high-frequency optoelectronic mixers (OEMs), which mix radiofrequency (RF) and optical signals. To date, 2D materials-based OEMs operating at 1.55 µm have only been demonstrated with graphene,[26,27,28,29] for which high mobility and efficient photodetection are mandatory properties to achieve high-performance.[28] For these OEMs, higher carrier mobility leads to higher conductivity and photoconductivity.[28] Compared to graphene, multilayer $PtSe_2$ has a lower carrier mobility but a higher IR absorption, which can be easily enhanced by increasing the number of layers.[12,30,31] Multilayer semimetallic $PtSe_2$ appears to be an excellent candidate to outperform the IR OEMs based on 2D materials. OEMs are key components, especially in telecommunications,[32] where signals with a high-frequency carrier need to be down-converted at the receiver. As data transmission increases rapidly, new wireless technologies such as 6G aim to increase signal bandwidths (> 0.5 GHz) and carrier frequencies (>10 GHz), requiring high-frequency OEMs.

The thickness-dependent electronic properties of $PtSe_2$ have been extensively investigated by theoretical[8,33,34,35] and experimental studies. Experimental observations have shown that $PtSe_2$ exhibits an indirect bandgap of ~ 2 eV for a monolayer (1 ML), which decreases to 0.6 - 1.1 eV for the bilayer,[35,36,37] and is semimetallic in the bulk.[9,38] However, the semiconductor-to-semimetal transition varies widely across different reported works, between 3 and 9 MLs.[13, 14,34,35,39,38,39,40,41] This could be related to the polycrystalline nature of grown films, which can present stacking polymorphism and Se vacancies, both influencing the band structure.[35,42,43] Indeed, the presence of both semiconducting and semimetallic phases has been reported in grown films.[30,31,44] Therefore, controlling the structural properties of $PtSe_2$ films is of foremost importance to understand and optimize their electronic properties.

Large research efforts have been devoted to its controlled synthesis using mechanical/liquid phase exfoliation,[17,38,45,46,47,48,49] VPT,[17,30,31,50] selenization of a pre-deposited platinum film (called TAC for "thermally assisted conversion") in a tube furnace[14,39,41,44,51,52,53,54] or in a molecular beam epitaxy (MBE) reactor,[15,55,56] and the "standard" MBE growth with simultaneous platinum (Pt) and selenium (Se) fluxes.[15,36,37,57,58] To date, the growth of high crystallinity $PtSe_2$ films on low-cost and insulating substrates, suitable for (opto)electronic devices, is still a challenge. To evaluate the crystalline quality, most groups[41,44,46,56,58] have measured the full width at half maximum of the $E_g$ Raman peak ($E_g$ FWHM): the narrower the $E_g$ peak, the higher the film crystallinity. Currently, $E_g$ FWHM < 5 $cm^{-1}$ is considered as the quality criterion for films grown by TAC,[44] with the lowest reported values ranging from 3.5 to 4.1 $cm^{-1}$ for polycrystalline films of 5.5 - 6.5 nm thickness (11 - 13 MLs).[41] MBE, which allows a precise control of the whole synthesis process, has enabled the growth of epitaxial $PtSe_2$ films on specific substrates such as Pt(111), bilayer graphene/6H-SiC(0001) and ZnO(0001).[15,55,57] Note that the $E_g$ FWHMs of these epitaxial films have not been reported. For all these studies, including epitaxial $PtSe_2$, the reported carrier mobilities remain considerably lower ($\leq$ 200 $cm^2.(V.s)^{-1}$)[15,39,40,41,44,50] than the predicted intrinsic mobility of $PtSe_2$ (> 4000 $cm^2.(V.s)^{-1}$). This indicates that further research efforts are required to achieve highly crystalline and homogeneous films on a large scale, in order to demonstrate high-performance (opto)electronic devices.

In this work, we first study the optical absorption at 1.55 µm and the sheet conductance of $PtSe_2$ as a function of the number of layers and show that thick ($\geq$ 10 MLs) films exhibit both high IR absorption and sheet conductance, *i.e.* the characteristics required for efficient OEMs. We secondly investigate the synthesis of thick (12 - 16 MLs) semimetallic $PtSe_2$ films by MBE on sapphire(0001) substrates. Sapphire was chosen because it is particularly well suited for optoelectronic applications: it is a low-cost, insulating and transparent substrate available in sizes up to 12 inches. The effects of the growth temperature, post-growth annealing and Se flux on the film crystallinity are studied by monitoring the two characteristic phonon modes of $PtSe_2$: the commonly used $E_g$ FWHM, and the $A_{1g}$ FWHM. Thirdly, in-depth structural analyses using X-ray diffraction (XRD) and scanning transmission electron microscopy (STEM) are performed to clarify the structural factors influencing $E_g$ and $A_{1g}$ FWHMs and to understand their impact on the electrical properties of $PtSe_2$ films. We show that a post-growth annealing process induces remarkable out-of-plane crystalline improvements, monitored by the $A_{1g}$ peak width, and leads to record sheet conductances for grown films, up to 1.6 mS. Compared to previous works[41,44,46,56,58] which only analyzed $E_g$ FWHM, the present study reveals the particular importance of also monitoring $A_{1g}$ FWHM: films with the lower $E_g$ and $A_{1g}$ FWHMs exhibit higher crystallinity in both in-plane and out-of-plane directions, as well as higher sheet conductances. Finally, we fabricate high-frequency 1.55 µm optoelectronic devices on a 2-inch sapphire substrate, based on high-quality 14 ML semimetallic $PtSe_2$ integrated into coplanar waveguides. We report photodetectors with a record bandwidth of 60 GHz and the first $PtSe_2$-based OEM with a bandwidth above 30 GHz.

## Results and Discussion

**1. Thickness Dependence of $PtSe_2$ Optical Absorption and Sheet Conductance.** In this section, we study the optical absorption and sheet conductance of $PtSe_2$ as a function of the thickness to determine the optimum film thickness for achieving high-performance 1.55 μm OEMs.

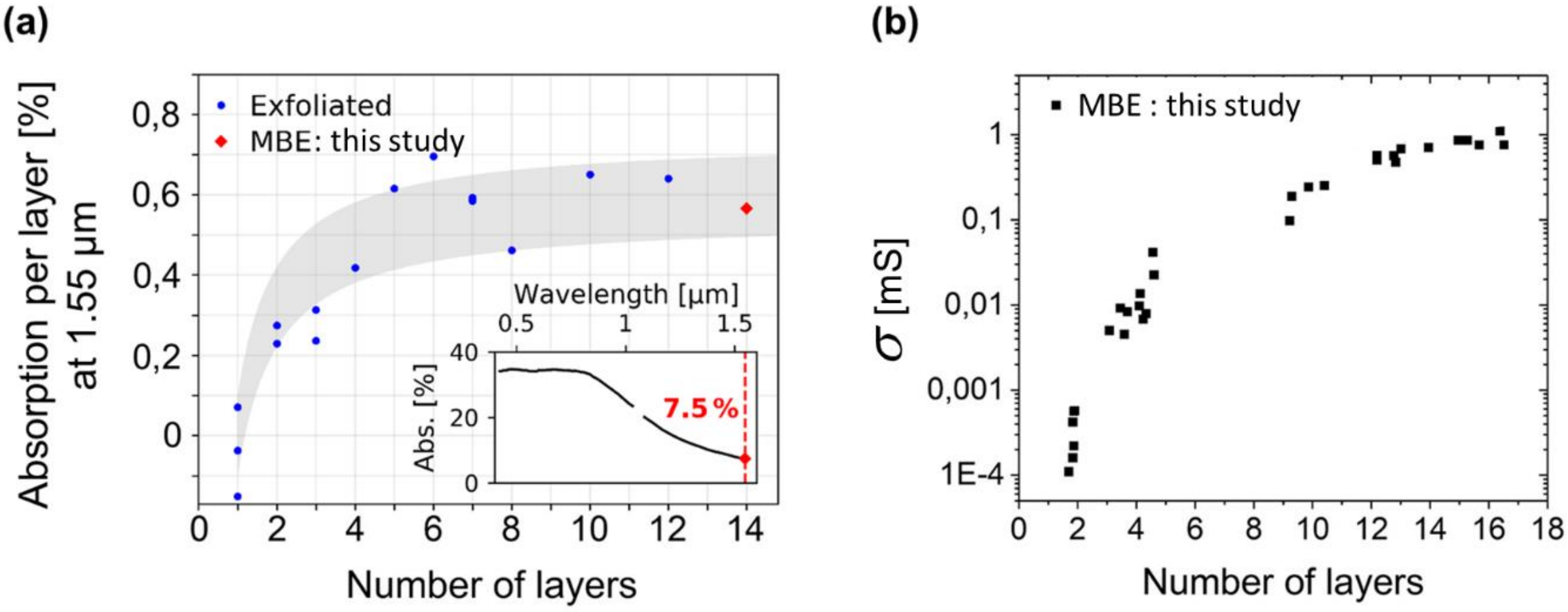

**Figure 1. Thickness dependence of optical and electrical properties of $PtSe_2$ films.** (a) Optical absorption per layer (at 1.55 μm) of $PtSe_2$ on a sapphire substrate as a function of the number of layers. The blue dots correspond to exfoliated $PtSe_2$ flakes and the red dot to a 14 ML $PtSe_2$ film synthesized by MBE in this study. The grey area is a guide for the eye. The inset shows the total optical absorption of this MBE film as a function of the optical wavelength. The dashed red line indicates an absorption of 7.5 % at 1.55 μm. (b) Sheet conductance of MBE-grown $PtSe_2$ films on a sapphire substrate as a function of the number of layers. The latter was determined by energy dispersive X-ray spectroscopy (EDX) measurements (see details in Figure S1).

First, optical absorption measurements were performed on $PtSe_2$ flakes exfoliated on quartz (experimental method described in Tharrault *et al.*[34]). Then, using a thin film optical model,[59] we calculated the optical absorption of the different flakes as if they were on a sapphire substrate instead of quartz. The blue dots in Figure 1a represent the optical absorption per layer of 1 to 12 MLs of $PtSe_2$ on sapphire. From 5 MLs, the absorption per layer of exfoliated flakes saturates at approximately 0.6 %, indicating that the absorption becomes proportional to the number of layers. This value is similar to that of the 14 ML $PtSe_2$ film synthesized by MBE (red dot in Figure 1a). The inset shows the total absorption of this MBE film as a function of the optical wavelength. The optical absorption, which is about 35 % in the visible region, decreases in the IR region but still reaches 7.5 % at 1.55 μm. This is a high value compared to a single layer of graphene on sapphire which absorbs only 1.2 % of the incident light.[59,60]

Secondly, the sheet conductance ($\sigma$) of $PtSe_2$ films synthesized by MBE in this study was investigated as a function of the number of layers (Figure 1b). $\sigma$ corresponds to the conductance of a squared-shaped $PtSe_2$ channel, *i.e.* the inverse of the sheet resistance ($R_{\square}$). Figure 1b shows that $\sigma$ is very low for thin (2 to 5 MLs) $PtSe_2$ films and increases from 0.1 to 1 mS between 9 and 16 MLs. The present study targets the fabrication of IR OEMs based on photoconductive $PtSe_2$. For single layer graphene, the photoconductivity scales with the conductivity.[28] Therefore, in order to fabricate efficient IR OEMs, we chose to focus on $PtSe_2$ films with thicknesses ranging from 12 to 16 MLs, which exhibit both high optical absorption at 1.55 μm and high sheet conductances.

**2. Optimization of $PtSe_2$ Synthesis.** We have investigated the MBE synthesis of $PtSe_2$ films on 20 x 20 $mm^2$ sapphire(0001) substrates (c-plane). During the synthesis, the heated substrate was exposed to Pt and Se fluxes of $\Phi(Pt) = 0.003$ Å.$s^{-1}$ and $\Phi(Se) = 0.2$ Å.$s^{-1}$. The film thickness was determined by energy dispersive X-ray spectroscopy (EDX) (see details in Figure S1). Raman spectroscopy ($\lambda$ = 532 nm) was used to characterize the crystalline quality of samples. The recorded spectra exhibit the two characteristic phonon modes of $PtSe_2$ (Figure 2a): the $E_g$ peak at ~ 178 $cm^{-1}$ and the $A_{1g}$ peak at ~ 206.5 $cm^{-1}$, corresponding respectively to the in-plane and the out-of-plane motion of Se atoms in opposite directions within a monolayer.[51] They also show a broad feature centered at ~ 235 $cm^{-1}$, attributed to the overlap of the two IR-active longitudinal optical (LO) modes named $A_{2u}$ and $E_u$.[51] Figure 2b presents the median values of the $E_g$ and $A_{1g}$ FWHMs as a function of the growth temperature (Tg), varying between 329 and 747 °C. The vertical bars correspond to the interquartile values of the FWHMs and represent the inhomogeneity of the $PtSe_2$ films (see Methods section). The lowest $E_g$ FWHMs (3.7 - 4.0 $cm^{-1}$), commonly attributed to the highest crystalline quality,[41,44,46] are obtained for samples grown at Tg between 463 and 570 °C. Above or below these temperatures, the $E_g$ FWHM increases above 4.5 $cm^{-1}$, indicating a deterioration in the film crystallinity. Note that the $A_{1g}$ FWHM follows exactly the same trend.

EDX measurements were performed on these films to extract the atomic ratio of Se/Pt (Figure 2c). Samples grown at Tg between 463 and 570 °C exhibit the highest ratios, ranging from 1.85 to 1.92. Such values smaller than 2 ($PtSe_2$ stoichiometry) are generally obtained on grown films.[17,41] Samples synthesized at Tg below 463 °C present slightly lower Se/Pt ratios (1.75 and 1.76), indicating higher Se deficiency. This ratio drops to 0.67 for the film grown at Tg = 597 °C and stabilizes at 0.4 around 700 °C. Note that at Tg = 597 °C, a strong dispersion of both $E_g$ and $A_{1g}$ FWHMs is observed (large vertical bars in Figure 2b), indicating a strong film inhomogeneity, as shown in Figure S2(c-d). These results demonstrate that the limiting growth temperature of $PtSe_2$ with $\Phi$(Se) = 0.2 Å.$s^{-1}$ lies between 570 and 597 °C. The abrupt decrease in the Se/Pt ratio for Tg ≥ 597 °C can be explained by the small Se sticking coefficient, which decreases exponentially with increasing temperature.[61] It could also be caused by a phase transition. Tong *et al.*[62] observed a phase change from $PtSe_2$ to $Pt_2Se$ alloy monolayer structure at temperatures above 600 °C. Therefore, Figure 2 demonstrates that high crystalline quality $PtSe_2$ films ($E_g$ FWHM $\leq$ 4 $cm^{-1}$ and $A_{1g}$ FWHM $\leq$ 4.5 $cm^{-1}$) can be synthesized using simultaneous Pt and Se fluxes ($\Phi(Se)/\Phi(Pt) \sim 70$) over a wide range of temperature (between 463 and 570 °C).

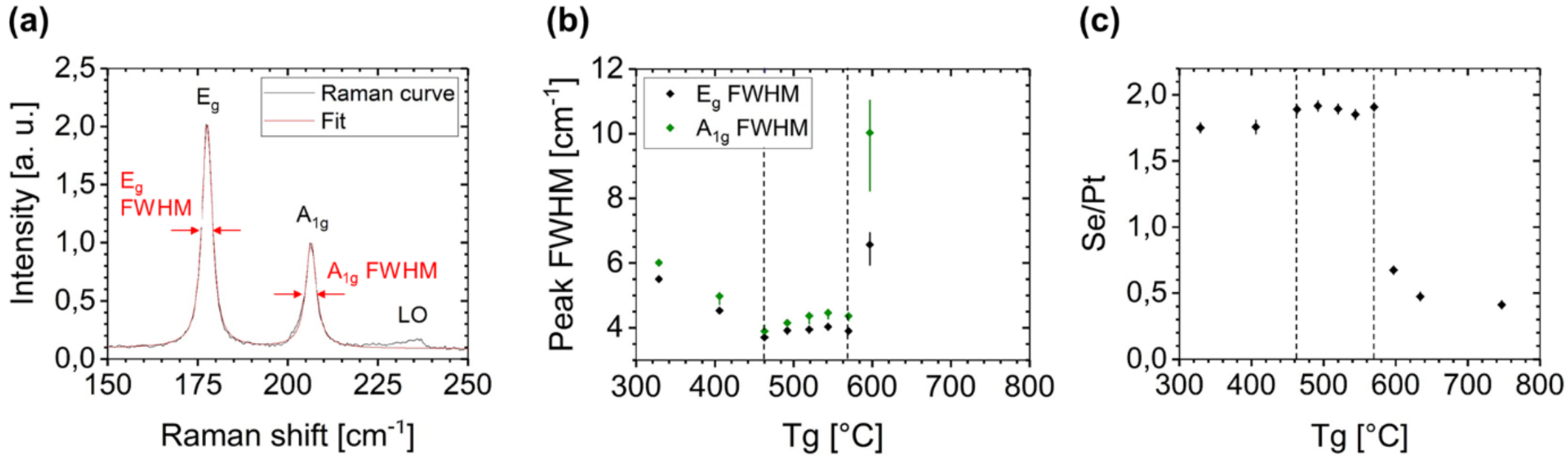

**Figure 2. Raman and EDX analyses of as-grown samples synthesized at different growth temperatures.** (a) Typical Raman spectrum ($\lambda$ = 532 nm) of a $PtSe_2$ film (black curve) showing the $E_g$ in-plane mode, the $A_{1g}$ out-of-plane mode and the LO modes. The intensity is normalized to the $A_{1g}$ peak. The $E_g$ and $A_{1g}$ peaks are fitted with Lorentzian functions (the red curve represents the sum of the two fits) to extract the FWHMs. (b) FWHMs of the $E_g$ peak (black dots) and the $A_{1g}$ peak (green dots) of $PtSe_2$ films synthesized at different growth temperatures (Tg) under a Se flux of 0.2 Å.$s^{-1}$. The dots are medians and the vertical bars correspond to interquartiles. From Tg = 634 °C, the Raman signal of $PtSe_2$ films becomes very weak (see Figure S2(a-b)) and it was not possible to extract accurate peak FWHMs. (c) Se/Pt atomic ratios of the corresponding samples measured by EDX. Here, the vertical bars correspond to the error on the Se/Pt values.

To further improve the crystalline quality of our $PtSe_2$ films, we have fixed the growth temperature at Tg = 520 °C and studied the effects of the Se flux and post-growth annealing on the Raman peaks. The Se flux was varied between 0.1 and 0.5 Å.$s^{-1}$ (the Pt flux was kept at 0.003 Å.$s^{-1}$). The annealing was then performed at Ta = 690 °C for 30 min under the same Se flux used during growth. The film thickness varies between 12 and 16 MLs. Note that we can neglect the influence of thickness on the $E_g$ and $A_{1g}$ FWHMs, since the films are more than 10 MLs thick, as demonstrated by Tharrault *et al*.[49]

Figure 3 compares the Raman data of $PtSe_2$ films grown without and with this post-growth annealing. Figure 3a shows $E_g$ FWHM as a function of the Se flux. Regardless of the Se flux used, annealed samples exhibit lower $E_g$ FWHMs than samples synthesized without annealing (as-grown samples). In addition, $E_g$ FWHM of the annealed samples slightly decreases with increasing Se flux: from 4.0 $cm^{-1}$ at 0.1 $Å.s^{-1}$ to only 3.6 - 3.7 $cm^{-1}$ at 0.5 $Å.s^{-1}$. Increasing the Se flux does not significantly modify the Se/Pt atomic ratio measured by EDX and there is also no obvious change in this ratio between as-grown and annealed samples (see Figure S3), demonstrating that no desorption of Se occurs during the post-growth annealing under Se flux. Comparing Figure 3b with Figure 3a shows that the main effect of the post-growth annealing is a strong reduction in the FWHM of the out-of-plane $A_{1g}$ mode, regardless of the Se flux. For example, at a Se flux of 0.5 $Å.s^{-1}$, the $A_{1g}$ FWHM drops from 4.3 - 4.7 $cm^{-1}$ to 3.5 - 3.8 $cm^{-1}$ due to the annealing. Therefore, by comparing the $E_g$ and $A_{1g}$ FWHMs of the as-grown and annealed samples, we demonstrate that the post-growth annealing at high temperature significantly improves the crystalline quality of $PtSe_2$ films. The lowest FWHMs of both peaks are obtained for the annealed samples synthesized under 0.5 $Å.s^{-1}$ of Se, indicating that the higher the Se flux, the higher the crystalline quality. The Raman mapping displayed in Figure S4 demonstrates the high homogeneity of these films at the micrometer scale.

Hilse *et al.*[58] have also studied the MBE synthesis of thick $PtSe_2$ films on a sapphire substrate and have obtained $E_g$ and $A_{1g}$ FWHMs of 5.2 and 6.5 $cm^{-1}$, respectively. Compared to their study, we have implemented higher $\Phi(Se)/\Phi(Pt)$ ratios (33 - 170 instead of 20), performed the growth at 520 °C instead of 200 °C and annealed the films at 690 °C instead of 400 °C. These elevated temperatures and Se fluxes result in as-grown and annealed samples exhibiting much lower FWHMs. Our annealed MBE samples synthesized under 0.5 $Å.s^{-1}$ of Se present $E_g$ and $A_{1g}$ FWHM down to ~ 3.5 $cm^{-1}$, approaching the values obtained with thick exfoliated $PtSe_2$ flakes, *i.e.* between 2.6 and 3.8 $cm^{-1}$ for the $E_g$ peak.[46,48,49]

Figure 3c compares the intensity ratio between the $A_{1g}$ and $E_g$ peaks ($I(A_{1g})/I(E_g)$) as a function of the number of layers. The increase in this ratio with increasing film thickness has been observed previously[41,46,49,51,57,58] and is explained by enhanced van der Waals (vdW) interlayer interactions.[51,57] It has been suggested that this ratio can serve as a robust metric for layer number across various $PtSe_2$ sample types.[46] Figure 3c shows that, for our 12 - 16 ML $PtSe_2$ films, the ratio is slightly affected by the film thickness, but is significantly influenced by our post-growth annealing at 690 °C. Annealed samples exhibit a much higher ratio (~ +40 %) compared to as-grown samples, as already observed by Hilse *et al*.[58] All these results tend to demonstrate that this ratio is influenced by both film thickness and crystalline quality.

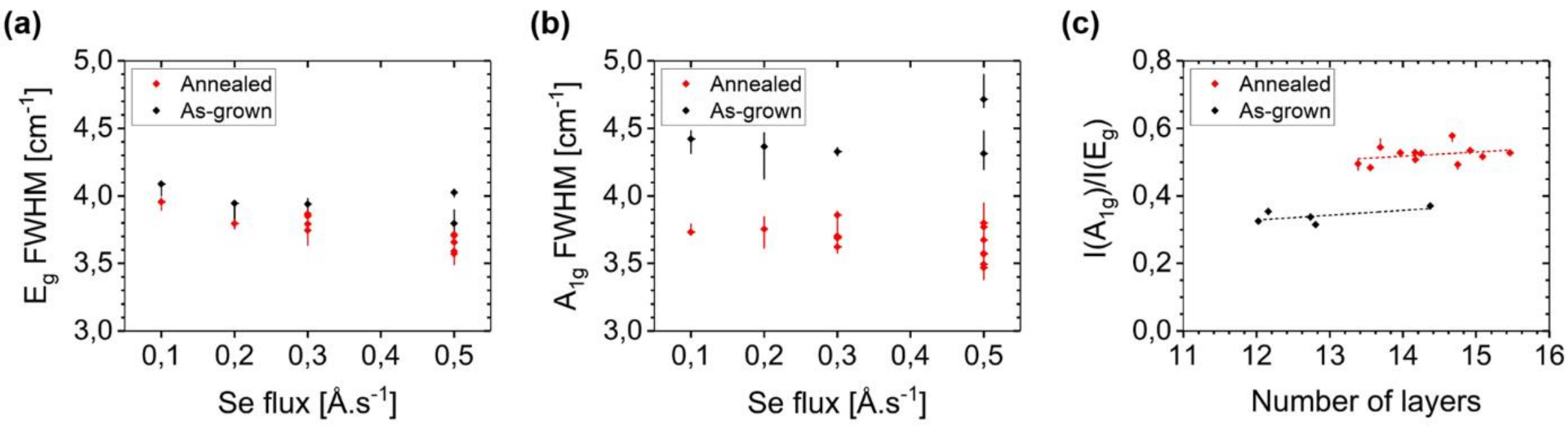

**Figure 3. Raman analysis ($\lambda$ = 532 nm) of as-grown and annealed films.** Samples are synthesized at 520 °C without (black dots) or with (red dots) a post-growth annealing at 690 °C for 30 min, under different Se fluxes. The dots are medians and the vertical bars correspond to interquartiles. 7 annealed samples were prepared under a Se flux of 0.5 Å.$s^{-1}$ to study the reproducibility of the synthesis. FWHMs of (a) the $E_g$ peak and (b) the $A_{1g}$ peak as a function of the Se flux. (c) Intensity ratio between the $A_{1g}$ and $E_g$ peaks ($I(A_{1g})/I(E_g)$) as a function of the number of layers. The latter was determined by EDX measurements (see details in Figure S1).

We also investigated the sheet conductance of these $PtSe_2$ films (Figure 4). We selected samples of similar thickness for this study: 12 - 13 MLs for as-grown samples and 14 - 15 MLs for annealed samples. Both as-grown and annealed samples exhibit purely ohmic behaviors. Figure 4a shows that $\sigma$ is similar for the as-grown samples and equal to 0.5 - 0.6 mS, regardless of the Se flux. In contrast, $\sigma$ of the annealed samples is much higher and increases with the Se flux (red dashed line): it goes from 1.17 mS for 0.1 $Å.s^{-1}$ to 1.36 - 1.59 mS for 0.5 $Å.s^{-1}$. Figure 4b shows that these measured sheet conductances appear to be well above the reported values ($\leq$ 0.3 mS) for grown $PtSe_2$ films, and that our record value of 1.6 mS is as high as the best value reported for exfoliated $PtSe_2$ of similar thickness (8 nm *i.e.* ~ 16 MLs).[38]

Figure 4c and 4d present $\sigma$ as a function of the $E_g$ FWHM and $A_{1g}$ FWHM, respectively. Regarding the annealed samples, $\sigma$ increases as the FWHM of both the $E_g$ and $A_{1g}$ peaks decreases, with a very similar slope. Figure 4c shows that as-grown and annealed films with similar $E_g$ FWHMs can exhibit different $\sigma$ values. In contrast, in Figure 4d we observe that the sheet conductance of the same films can be fitted as a linear function of $A_{1g}$ FWHM. Finally, Figure 4 clearly demonstrates that the post-growth annealing at high temperature significantly increases the sheet conductance of $PtSe_2$ films, and that this improvement is mainly exhibited by the FWHM of the out-of-plane $A_{1g}$ mode. Our annealed MBE samples synthesized under 0.5 $Å.s^{-1}$ of Se exhibit the highest sheet conductances (1.36 - 1.59 mS). Thus, our optimum growth conditions are determined to be Tg = 520 °C, Ta = 690 °C and $\Phi(Se)/\Phi(Pt) = 170$.

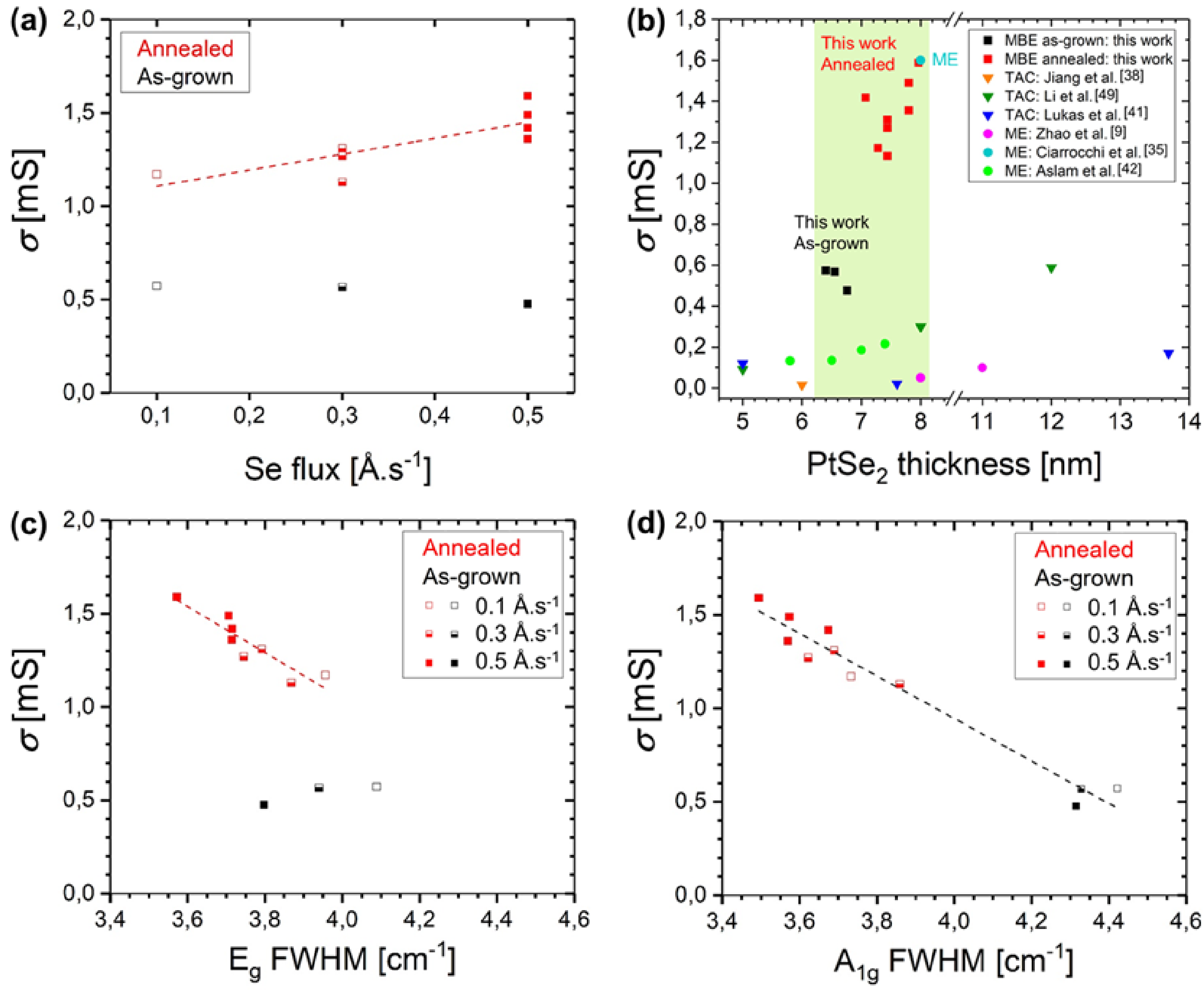

**Figure 4. Sheet conductance of $PtSe_2$ films.** Samples (12 - 15 MLs) are synthesized at 520 °C without (black dots) or with (red dots) a post-growth annealing at 690 °C for 30 min, under different Se fluxes. $\sigma$ is plotted as a function of (a) the Se flux, (c) the $E_g$ FWHM and (d) the $A_{1g}$ FWHM. Empty, half-empty and full squares correspond to Se fluxes of 0.1, 0.3 and 0.5 $Å.s^{-1}$, respectively. (b) displays the state-of-the-art sheet conductances of $PtSe_2$ films as a function of their thickness. The samples were synthesized using different methods: mechanical exfoliation (ME) of a bulk crystal, TAC and MBE in this study. Note that the highest sheet conductance of 1.6 mS found in the literature (cyan dot) was measured at 10 K.

The air stability of thick $PtSe_2$ films was investigated using Raman spectroscopy, EDX and sheet conductance measurements. Figure S5a shows the Raman spectra recorded on an annealed sample a few days after the synthesis and after 1.5 years in air. The spectra are similar, as well as the medians of both the $E_g$ and $A_{1g}$ FWHMs. There is also no significant change in the Se/Pt ratio after 1.5 years exposure to air (Figure S5(b-c)). Finally, $\sigma$ of the film a few days after the synthesis and after 1.5 years in air are 1.49 mS and 1.46 mS, respectively. Therefore, the crystalline quality and sheet conductance of multilayer $PtSe_2$ are preserved after 1.5 years in air, demonstrating the excellent stability of this 2D material and its great potential for device applications.

**3. XRD and STEM Structural Characterizations on As-Grown and Annealed Samples.** In-depth structural analyses were carried out on $PtSe_2$ films synthesized at 520 °C without and with a post-growth annealing at 690 °C under 0.5 $Å.s^{-1}$ of Se (as-grown and annealed films, respectively). XRD and STEM techniques are implemented to determine the structural changes induced by the post-growth annealing, which drastically reduces $A_{1g}$ FWHM and increases sheet conductance.

First, $PtSe_2$/sapphire(0001) samples were characterized at the macroscopic scale by grazing incidence X-ray diffraction (GIXRD) measurements. Bragg scans along the in-plane $Al_2O_3$<11-20> direction were performed on both films. Figure 5a shows that only crystallographic planes with Miller index l = 0 are observed, implying that the two films are fiber textured with c-planes parallel to the substrate surface. The azimuthal φ scan (Figure S6) across the $PtSe_2$[11-20] direction confirms the fiber texture with domains exhibiting random in-plane orientations. Bragg scans also provide useful information on the in-plane lattice parameters and domain sizes. Considering a hexagonal crystal symmetry, the in-plane lattice parameters "a" are calculated to be equal to 3.698 ± 0.002 Å and 3.716 ± 0.001 Å for the as-grown and annealed samples, respectively. Thus, the post-growth annealing induces an increase in the lattice parameter, approaching the experimental value of 3.727 Å measured for monocrystalline $PtSe_2$ bulk.[63] We implemented the Williamson-Hall method[64] to evaluate the contribution of the in-plane domain size ($D_{WH}$) and the lattice microstrain ($\varepsilon$) to the broadening of the diffraction peaks (see details in Table S1 and S2). According to this method:

$$\Delta(2\theta_\chi)\cos(\theta_\chi) = \frac{K\lambda}{D_{WH}} + 4\varepsilon\sin(\theta_\chi)$$

where $2\theta_\chi$ is the diffraction angle, $\Delta(2\theta_\chi)$ is the diffraction peak FWHM deconvoluted from the instrument width, K is the Scherrer constant (0.94 for isotropic crystallites) and $\lambda$ is the X-ray wavelength (1.54056 Å).

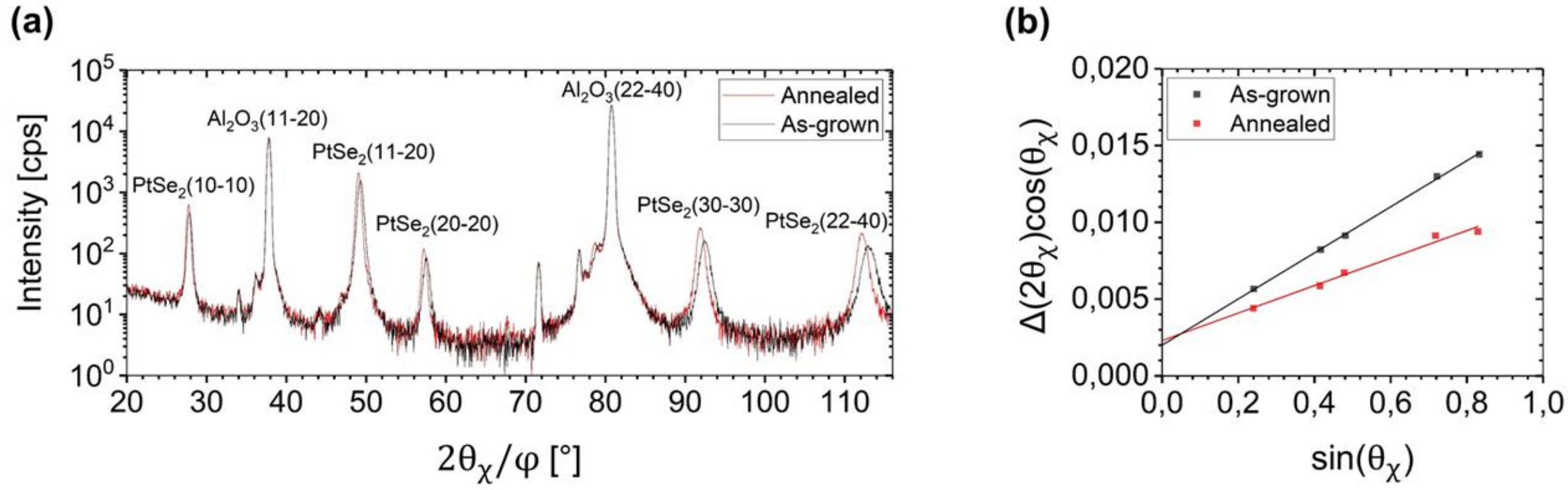

**Figure 5. GIXRD measurements.** (a) GIXRD Bragg scans of the as-grown (black curve) and annealed (red curve) $PtSe_2$ films along the in-plane $Al_2O_3$<11-20> direction. The FWHM of the five $PtSe_2$(hki0) diffraction peaks ($\Delta(2\theta_\chi)_{meas}$) are extracted by fitting them with PearsonVII functions (Table S1). (b) Calculation of the in-plane domain size ($D_{WH}$) and lattice microstrain ($\varepsilon$) using a Williamson-Hall plot applied to these five diffraction peaks (see details in Table S1 and S2).

In Figure 5b, $\Delta(2\theta_\chi)\cos(\theta_\chi)$ is plotted as a function of $\sin(\theta_\chi)$ and fitted with a linear function. For both the as-grown and annealed samples, the experimental data are well fitted by lines, as predicted by the method. From the slope and y-intercept, we derive $D_{WH}$ and $\varepsilon$ of 724 ± 54 Å and 0.38 % for the as-grown film and 630 ± 107 Å and 0.22 % for the annealed film (Table S2). Thus, the in-plane domain size is approximately the same for the as-grown and annealed samples and the main effect of the post-growth annealing is an important reduction in the non-uniform in-plane lattice distortions, by a factor of almost 2. The observed local microstrains can be due to point defects and/or dislocations, suggesting that the defect density is lower in the annealed sample than in the as-grown sample.[65]

Figure 6 shows the high-resolution X-ray diffraction (HRXRD) 2θ/ω Bragg scans collected on the as-grown and annealed samples. As expected, only the $PtSe_2$ diffraction peaks corresponding to crystallographic planes with Miller indices h = k = 0 are detected. The $PtSe_2$ diffraction peaks, as well as the fringes around the $PtSe_2$(0001) peak are better defined for the annealed sample compared to the as-grown sample. According to the positions of the diffraction peaks, the out-of-plane lattice parameter "c" is smaller for the annealed sample compared to the as-grown sample, 5.165 ± 0.02 Å with respect to 5.300 ± 0.10 Å, and approaches the value of monocrystalline $PtSe_2$ bulk (5.07 Å).[63] The asymmetric shape of the diffraction peaks is more pronounced in the as-grown sample, with a tail towards low angles, revealing a gradual distribution of the lattice parameter "c" towards higher values. Therefore, HRXRD analysis shows that the annealed sample presents a higher out-of-plane crystalline quality, with a smaller and more uniform interlayer distance compared to the as-grown sample.

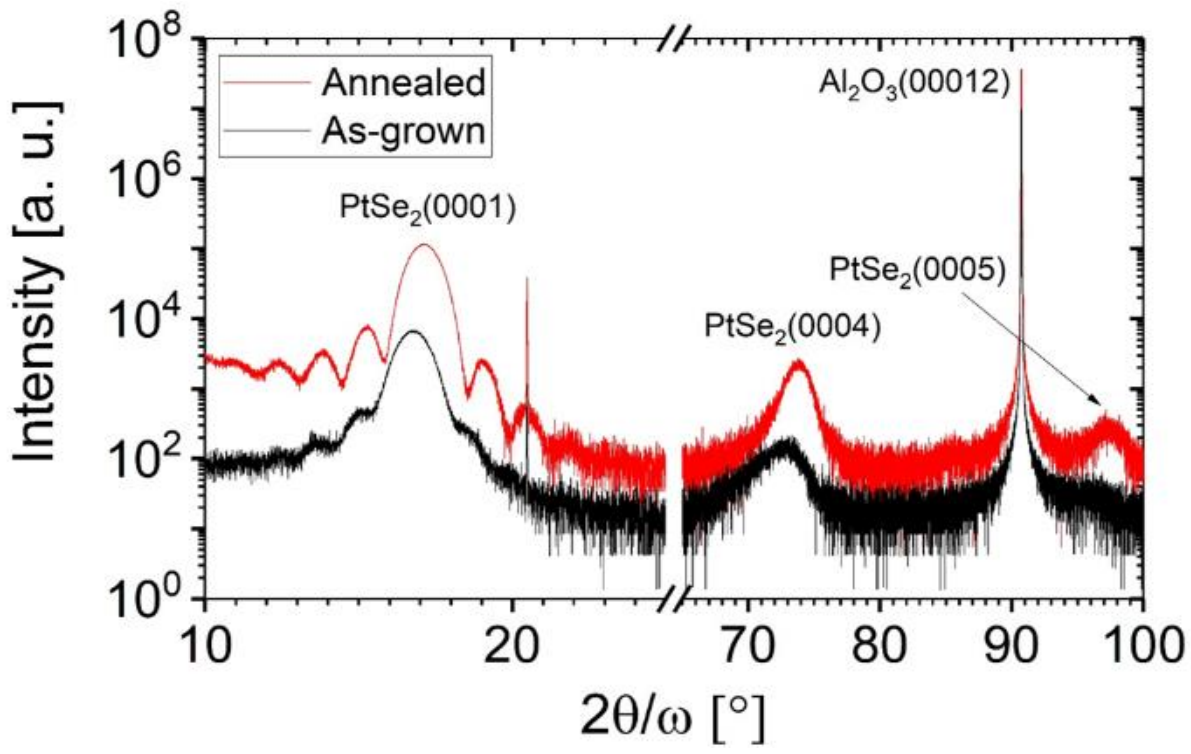

**Figure 6. HRXRD measurements.** HRXRD Bragg scans of the as-grown (black curve) and annealed (red curve) $PtSe_2$ films along the $Al_2O_3$<0001> direction.

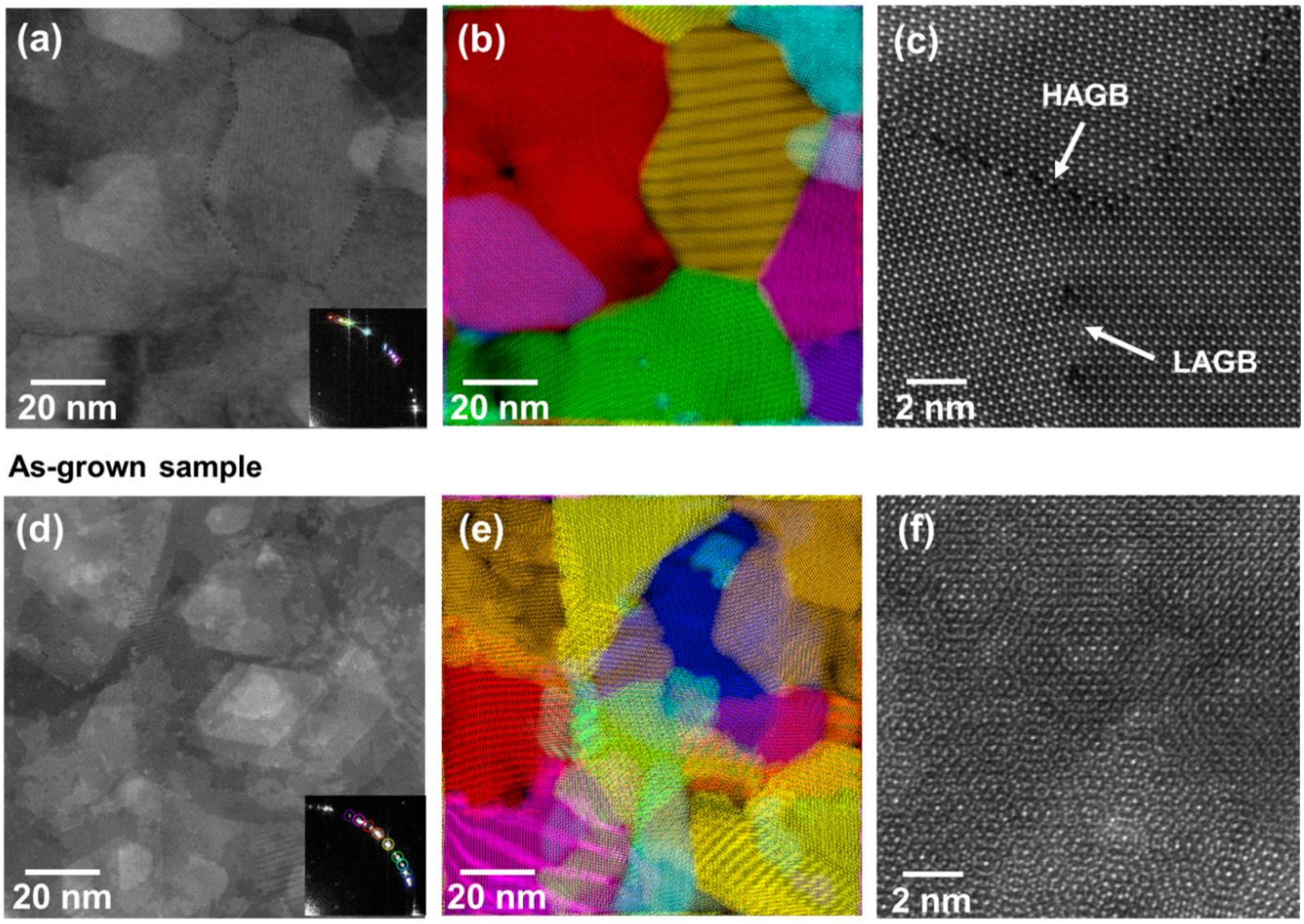

**Figure 7. High resolution STEM images.** (a) HAADF-STEM image of the annealed sample with (b) the corresponding FFT-based composite DF image. (c) Typical HAGB and dislocations aligned as LAGB. (d) HAADF-STEM image of the as-grown sample with (e) the corresponding FFT-based composite DF image and (f) the typical moiré structures. The color code used for the DF images is given in the FFT in the inset of their original HAADF images.

The detailed structural properties of the $PtSe_2$ films were investigated using STEM. Figure 7a shows a plan-view high angle annular dark field (HAADF) image of the annealed film. Figure 7b is a composite dark field (DF) image reconstructed using the Fast Fourier Transform (FFT), where each color represents an in-plane orientation indicated in the inset of Figure 7a. Atomic resolution analysis (Figure 7c) allows identifying highly crystalline domains exhibiting random in-plane orientations. These are laterally connected either by continuous line defects, the so-called high angle grain boundaries (HAGBs), or by isolated dislocation cores defining low angle grain boundaries (LAGBs) (see detailed GB structures in Figure S7). Considering the $PtSe_2$ film thickness (~ 14 MLs), the perfect vertical alignment of the crystal lattice as well as the GBs and dislocations within the film thickness demonstrates a single crystalline character along the c-axis. STEM image analyses (Figure 7b and Figure S8) revealed that most of crystalline domains have in-plane sizes falling within the range of 50 - 80 nm. In the as-grown sample, domain boundaries are not well aligned along the c-axis, as observed in the HAADF image (Figure 7d). The composite DF image (Figure 7e) and the atomic resolution image (Figure 7f) suggest that several domains with different in-plane orientations are partially superimposed, resulting in complex moiré structures. The domain sizes, as defined by the DF images for specific in-plane orientations, are also found to be approximately in the range of 50 - 80 nm (Figure S8). This indicates that the annealing process induces no obvious change in the in-plane domain size, in good agreement with the GIXRD measurements. In contrast to the annealed sample, slight variations in the local in-plane orientation can often be observed within the identified individual domains of the as-grown film, suggesting a relatively low crystallinity, with the presence of point defects and dislocations (see detailed domain analysis of the as-grown sample in Figure S8).

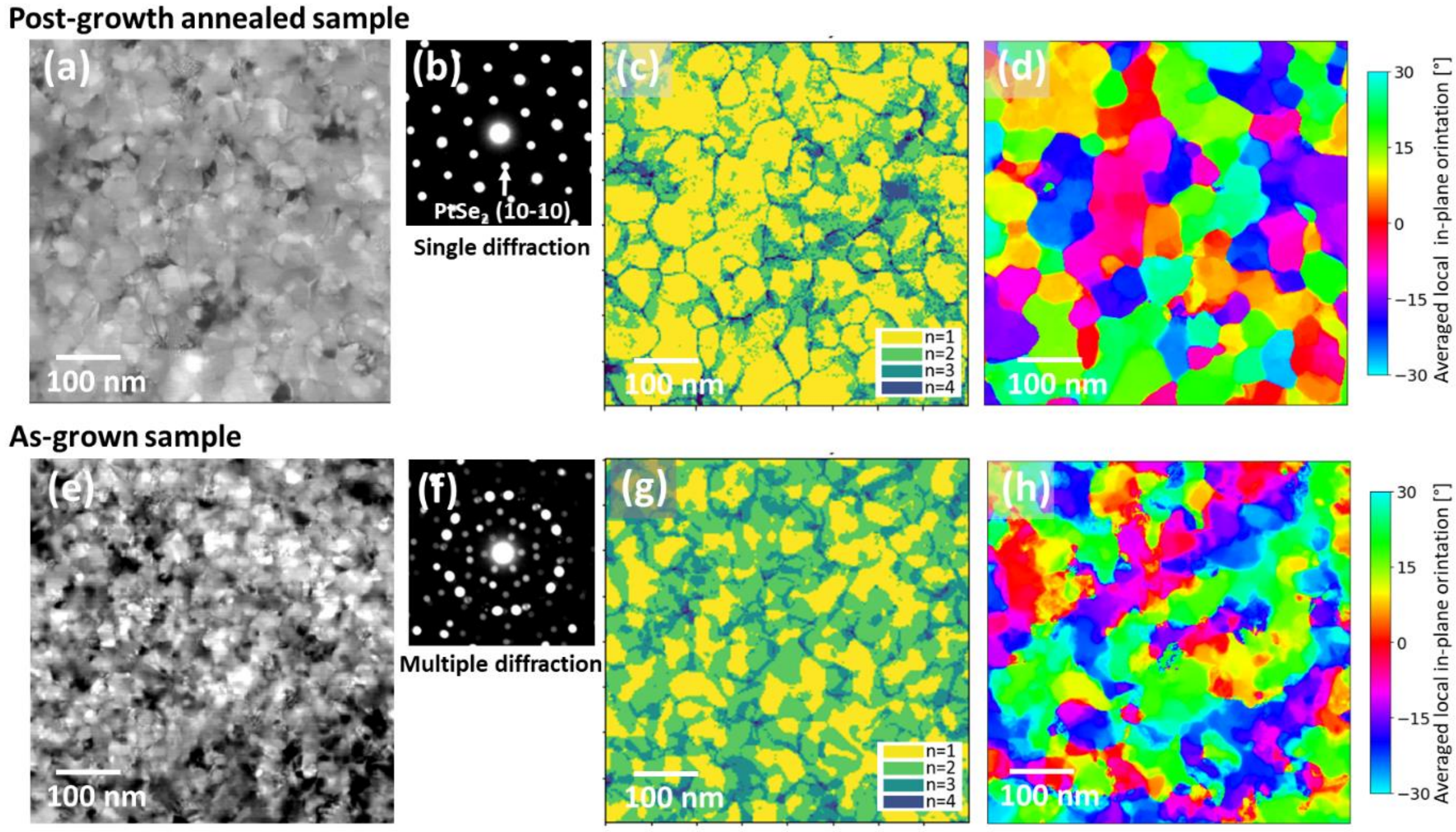

**Figure 8. Structural maps reconstructed from the 4D diffraction datasets of the annealed and as-grown samples.** (a, e) Virtual ADF images. Typical diffraction patterns observed on (b) the annealed and (f) as-grown samples. (c, g) Maps of the number of hexagonal symmetries at each pixel position, corresponding to the number of locally superimposed twisted domains ($n$), with the color codes in the inset. (d, h) Locally averaged in-plane orientation maps.

We have then carried out 4D-STEM analyses to investigate the large-scale distribution of the crystalline domains in the $PtSe_2$ films. An electron diffraction pattern was recorded at each pixel of the electron beam scan (400 x 400 pixels) over sub-micron areas (~ 570 x 570 nm$^2$). Figure 8a and 8e display virtual annular dark field (ADF) images reconstructed from the 4D diffraction datasets, as well as typical diffraction patterns (Figure 8b and 8f) of the annealed and as-grown samples, respectively. To assess the stacking quality of the multilayer $PtSe_2$ films, the number of hexagonal symmetries observed in individual diffraction patterns was used to identify the superposition of twisted domains at each position in real space. The resulting maps are shown in Figure 8c and 8g, where the color code in the inset directly represents the number of hexagonal symmetries, corresponding to the number of superimposed twisted domains along the film thickness ($n$). In-plane orientation maps (Figure 8d and 8h) are also reconstructed by measuring the direction of the hexagonal symmetry in the diffraction pattern recorded for each position.

Figure 8c demonstrates that the annealed sample exhibits fairly well vertically aligned domains ($n = 1$) over a large area, with a few misaligned regions ($n \geq 2$) at domain junctions. As most of the domains present a single in-plane orientation within the depth, (with a single diffraction pattern, as shown in Figure 8b), the individual domains are relatively well defined by the color codes of the averaged in-plane orientation map in Figure 8d. These maps show a film composed of vertically single-crystalline domains, laterally connected with random in-plane orientations. For the as-grown film, smaller islands with a single in-plane orientation ($n = 1$) are observed, surrounded by regions consisting of a superposition of multiple domains (Figure 8g). The orientation map in Figure 8h represents the average of the multiple in-plane orientations present within the film thickness and, therefore, does not reflect the individual crystalline domains. To further resolve the individual crystalline domains in the film, defined by the local in-plane orientation, dark-field (DF) imaging was performed on the same 4D diffraction datasets for both samples (Figure S9 and S10). This observation confirms our previous findings, indicating no obvious difference in the in-plane crystalline domain size between the two samples. However, the as-grown film exhibits extensive domain superposition at the in-plane grain boundaries, resulting in a much lower stacking quality compared to the annealed film.

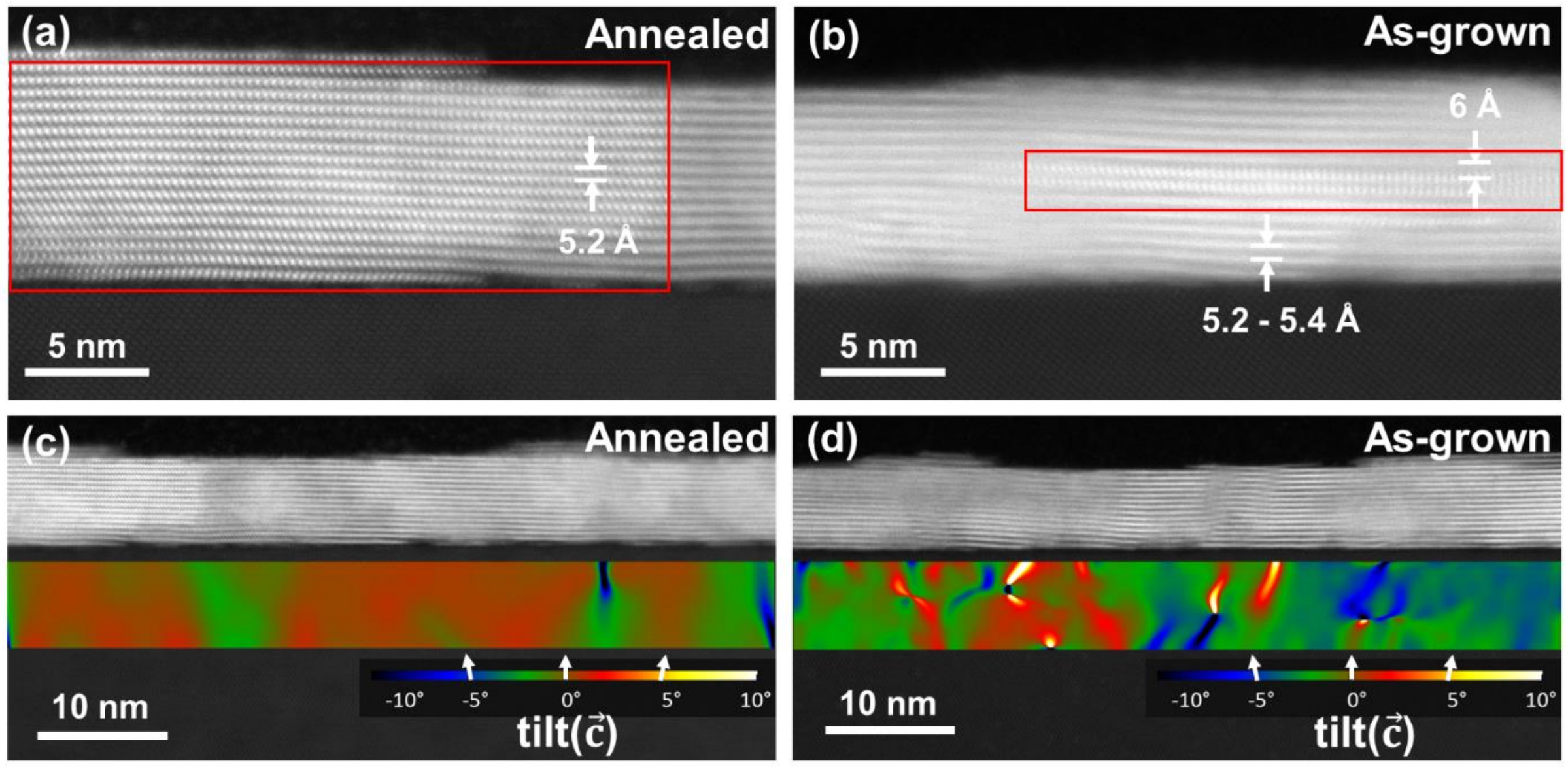

**Figure 9. Cross-sectional HAADF-STEM images.** Images of (a) the annealed and (b) as-grown samples, with single domains oriented in a zone axis indicated by red squares. Typical film morphologies of (c) the annealed and (d) as-grown samples. The local tilt angle between $PtSe_2$ and sapphire c-planes, tilt($\vec{c}$), was determined by GPA using the $PtSe_2$(0001) spot. The corresponding tilt($\vec{c}$) map is displayed below each cross-sectional image. Intense vertical lines appear due to the interface effects.

Cross-sectional STEM analyses were carried out to investigate the out-of-plane morphology of the films. Figure 9a and 9b show atomic resolution HAADF-STEM images of typical structures investigated in the annealed and as-grown samples, respectively. In the annealed film, each $PtSe_2$ layer is horizontally connected to those in neighboring domains. Each single domain consists of flat $PtSe_2$ layers parallel to each other, with an average interlayer distance of 5.2 Å, with quite small variations. In Figure 9a, a crystalline domain highlighted by a zone axis (surrounded by the red square) exhibits vertically single crystallinity and the domain thickness corresponds to the film thickness. This confirms a two-dimensional arrangement of highly crystalline domains with a perfect horizontal conjunction. In the as-grown film, each $PtSe_2$ layer is also horizontally connected, forming a continuous film. However, ruptures in the crystal structure within the film thickness are often identified by increased interlayer distances that we assume to be boundaries between superimposed twisted domains. This is confirmed in Figure 9b where a single crystal domain composed of 2 MLs oriented in a zone axis (surrounded by the red square) is separated from the upper and lower domains by an in-plane misorientation and an increased interlayer distance. The interlayer distance varies between 5.2 and 5.4 Å within areas considered as an individual domain, indicating a relatively low crystallinity, and increases to up to 6 Å between domains. These results clearly show the stronger out-of-plane disorder in the as-grown sample, with a higher and non-uniform interlayer distance value, compared to the annealed sample. This is in good agreement with the HRXRD 2θ/ω Bragg scan showing asymmetric peak shapes shifted towards low angles (*i.e.* high interlayer distance) for the as-grown sample (Figure 6).

Cross-sectional STEM images (Figure 9c and 9d) are also used to extract information on the local out-of-plane misorientation. The local tilt angle of the $PtSe_2$ c-plane relative to the sapphire c-planes, tilt($\vec{c}$), is determined by FFT-based Geometric Phase Analysis (GPA) using the $PtSe_2$(0001) spot.[66,67] This allows to generate orientation field maps, which are shown below each cross-sectional image in Figure 9c and 9d. In the annealed film, the variation in tilt($\vec{c}$) is mainly observed along the in-plane direction, whereas in the as-grown sample, it varies in the volume of the film, along both the in-plane and out-of-plane directions. For an effective and quantitative comparison of out-of-plane tilt disorders between different samples, we determined characteristic standard deviation values of the local c-plane tilt angle, SD tilt($\vec{c}$), over a large area: tilt($\vec{c}$) values are collected on all pixels of orientation field maps generated over a total width of 600 nm (see details in Figure S11). The characteristic SD tilt($\vec{c}$) are 0.92 ° for the annealed sample and 1.65° for the as-grown sample. The SD tilt($\vec{c}$) measured from the orientation field maps reflects several structural factors concerning the planarity of the film and its component layers: domain tilt, interlayer parallelism and individual layer flatness. Although the cross-sectional STEM provides a rather local analysis, the characteristic SD tilt($\vec{c}$) helps evaluating the out-of-plane crystalline quality of the films and reveals a clear improvement in the planarity factors in the annealed sample compared to the as-grown sample.

From our observations, we conclude that a three-dimensional (in-plane and out-of-plane) recrystallization is induced by the post-growth annealing at high temperature. In particular, the annealing process leads to a drastic improvement in the out-of-plane crystallinity. Moreover, a significant reduction in the defect density and a rearrangement of GBs are also revealed by the GIXRD and STEM analyses. Contrary to 3D bulk systems, the migration of dislocations and GBs in 2D materials takes place through the in-plane displacement of a few atoms around the dislocation cores.[68] Such phenomena induced by a thermal treatment have already been reported in monolayer graphene and few-layer TMDs, which lead to the rearrangement of dislocations and GBs for lower energy configurations,[69,70] and also enable the healing of rotational disorders and stacking faults when the binding energy facilitates it.[71,72,73] We assume that during the post-growth annealing, the high-energy configuration of the as-grown films, characterized by a high density of defects and strong microstrain, triggered in-plane defect migrations. This enabled the effective thermal healing of multiple types of structural defects within the multilayer films, lowering the overall system energy.[74,75] For instance, the disordered dislocations and GBs spread throughout the as-grown film are reorganized into horizontally well-ordered and vertically aligned HAGBs and LAGBs. This GB migration also reflects the relaxation of the system towards a lower-energy configuration, which generally enhances binding energy.[43,72] Here, we observe a reduction in out-of-plane tilt disorder, a decrease in interlayer distance and an improved vertical alignment of domains at GBs. These observations emphasize the important role of the interlayer interaction in such structural modification of multilayer films. It is worth noting that this in-plane crystal reorganization is spatially limited as no obvious difference in the in-plane domain size was observed between as-grown and annealed samples. In contrast, the vertical alignment of the in-plane domain boundaries results in a significant improvement in the vertically single crystalline character of the $PtSe_2$ films.

**4. $E_g$ and $A_{1g}$ FWHMs as Metrics for Crystalline Quality and Sheet Conductance.** Raman spectroscopy appears to be an effective technique for assessing the crystalline quality and sheet conductance of $PtSe_2$ samples. In addition, this fast and non-destructive characterization method is well suited for systematic analysis. For samples synthesized by TAC, Lukas *et al.*[44] have shown that the $E_g$ FWHM correlates with the in-plane domain size and the $R_{\square}$. In section 2 of the present paper, we have demonstrated for the first time the particular importance of the $A_{1g}$ FWHM to evaluate both the crystalline quality and the sheet conductance of thick $PtSe_2$ films. In-depth XRD and STEM results presented in section 3 can be used to clarify the correlations between the Raman peak FWHMs and the structural properties of the multilayer $PtSe_2$ films, and also between these structural properties and the sheet conductance. Figure 10a displays the $E_g$ and $A_{1g}$ FWHMs of two thick $PtSe_2$ films synthesized by TAC[44] together with our as-grown and annealed MBE films. It also provides their sheet conductances and key structural information: in-plane domain sizes, SD tilt($\vec{c}$) values, as well as schematics of the domain arrangement within the film thickness. A monocrystalline $PtSe_2$ flake (> 10 MLs) obtained by mechanical exfoliation is also included as a reference sample.[49]

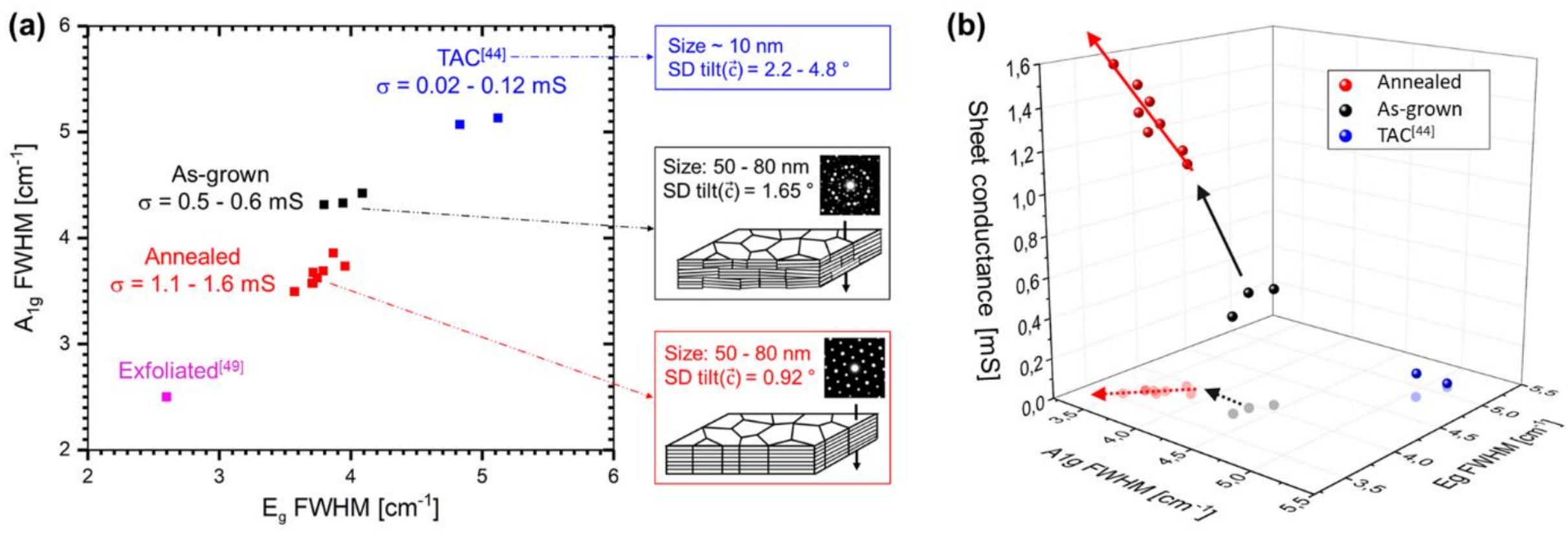

**Figure 10. Correlation between Raman metrics, crystallinity and sheet conductance of thick $PtSe_2$ films.** (a) 2D plot for thick (10 - 15 MLs) $PtSe_2$ films presenting their crystalline quality and sheet conductance as a function of their Raman peak FWHMs. It displays the $E_g$ and $A_{1g}$ FWHMs of two $PtSe_2/SiO_2/Si$ films synthesized by TAC (blue dots) with thicknesses of 5 and 7.6 nm (10 - 15 MLs range),[44] of our as-grown (black dots) and annealed (red dots) $PtSe_2$/sapphire(0001) films (12 - 15 MLs) synthesized by MBE under different Se fluxes (0.1 - 0.5 Å.s$^{-1}$) and of the thick (> 10 MLs) exfoliated $PtSe_2$ flake (purple dot) presenting the lowest reported $E_g$ and $A_{1g}$ FWHMs[49] (the sheet conductance was not reported). It also shows the in-plane domain size and SD tilt($\vec{c}$) values extracted from STEM analyses, and schematics of the domain arrangement within the film thickness with the characteristic diffraction patterns of the MBE films. (b) 3D plot showing the sheet conductance of the same $PtSe_2$ samples as a function of $E_g$ and $A_{1g}$ FWHMs. The arrows are guides for the eyes.

First, we investigate the correlations between $E_g$ and $A_{1g}$ FWHMs and the crystallinity of samples. It is worth noting that MBE films have much larger in-plane domain sizes and much lower $E_g$ FWHM compared to TAC films, while all MBE films exhibit similar $E_g$ FWHM and in-plane domain sizes. This confirms that the $E_g$ FWHM is well correlated with the in-plane domain size, which is one of the known dominant factors contributing to the in-plane crystalline quality.[44] Note that the slight decrease in $E_g$ FWHM between the as-grown and annealed MBE samples may be attributed to the significant reduction in defect density investigated by both XRD and STEM analyses. In contrast, the main difference between the annealed and as-grown MBE samples observed by Raman spectroscopy is an important reduction in the $A_{1g}$ FWHM for annealed samples. This can be related to the significant improvements in the out-of-plane crystallinity revealed by the in-depth structural analyses in section 3; in particular the increase in domain thickness and the reduction in out-of-plane tilt disorder (SD tilt($\vec{c}$) of 0.92 ° compared to 1.65 °). The strong reduction in both $A_{1g}$ FWHM and SD tilt($\vec{c}$) in as-grown films compared to TAC samples is also consistent with a higher out-of-plane crystallinity. These results suggest that $E_g$ and $A_{1g}$ FWHMs predominantly reflect the in-plane and out-of-plane crystalline quality, respectively. Therefore, these two Raman peak widths need to be investigated together and are relevant metrics to evaluate the overall crystallinity of $PtSe_2$. This assumption is supported by the thick (> 10 MLs) $PtSe_2$ flake exfoliated from a single crystal,[49] which exhibits significantly lower $E_g$ and $A_{1g}$ FWHMs of 2.6 and 2.5 $cm^{-1}$, respectively. Note that for other TMDs, such as $MoS_2$ and $MoSe_2$, it has already been demonstrated that the lower the $A_{1g}$ FWHM, the higher the crystalline quality of multilayer films.[76,77,78,79]

Secondly, we analyze the structural parameters that can influence the sheet conductance. As-grown films exhibit higher $\sigma$ than TAC samples: 0.5 - 0.6 mS compared to 0.02 - 0.12 mS. This can be explained by larger in-plane domain sizes (50 - 80 nm instead of 10 nm) constituting the films, implying a lower GB density. It is interesting to note that not only $E_g$ FWHM but also $A_{1g}$ FWHM is smaller for the as-grown films. This suggests the influence of the out-of-plane crystalline quality on the sheet conductance. Among others, the improved film flatness, as evaluated by the SD tilt($\vec{c}$) values (reduced to 1.65 ° compared to 2.2 - 4.8 °), might be one of the structural parameters that increase the sheet conductance. The MBE samples exhibit similar in-plane domain sizes, but the sheet conductance of annealed films is significantly higher than that of as-grown films (1.1 - 1.6 mS compared to 0.5 - 0.6 mS). This can be explained by their different out-of-plane structural properties, which are monitored by the $A_{1g}$ FWHM. The domains constituting the annealed films have typically a thickness equal to the film thickness (14 - 15 MLs) and thus, exhibit a metallic behavior.[17,38,39,41,50] In contrast, we often observed the superposition of 2 to 4 twisted domains within the film thickness of as-grown samples. These domains which are thinner than the film thickness, as schematized in Figure 10a, can exhibit locally semiconducting behaviors,[13,15,17,38,39,40,41,45,50] with various electronic properties depending on their local thickness.[11,33,34] Therefore, we can assume that as-grown films consist of a mixture of semiconducting and semimetallic phases, as observed in other studies.[30,44,31] A much lower local conductance is expected for semiconducting phases compared to semimetallic ones.[12,38,41,45] In addition, higher contact resistances are also expected between semiconducting domains with different electronic properties and also between semimetallic and semiconducting domains.[46] Therefore, achieving a domain thickness equal to the film thickness is particularly important to reach high sheet conductance. Annealed films, which satisfy this criterion, exhibit record sheet conductances for grown films, as shown in Figure 4b. If this condition is not satisfied, the sheet conductance strongly depends on the thickness of component domains. We propose a new criterion for high quality thick $PtSe_2$ films ($\geq$ 10 MLs), which is both $E_g$ and $A_{1g}$ FWHMs $\leq$ 4 $cm^{-1}$.

The impact of the post-growth annealing on the $PtSe_2$ sheet conductance might arise from an original mechanism: in conventional materials, the increase in crystallinity reduces carrier scattering, thereby increasing electronic mobility and sheet conductance. In contrast, in multilayer $PtSe_2$, the increased out-of-plane crystallinity can also change the nature of the material itself, by affecting its local electronic structures, turning it from a semiconductor to a semimetal. This observation may also explain the large variation in the number of layers for which the semiconducting-to-metallic transition occurs reported in different works.[17,13,34,35,38,39,40,41,43] Note that Figure 1b, which plots the sheet conductance of only as-grown films (with unknown domain thickness) as a function of the film thickness, cannot be used to study this transition. For an accurate determination of the layer number for which this transition happens, it is essential to ensure the vertical alignment of component domains.[34]

Figure 10b plots in 3D the sheet conductance of the same samples as a function of both the $E_g$ and $A_{1g}$ FWHMs. It clearly shows the correlation between these Raman metrics and the sheet conductance: films with the lower FWHMs of both the $E_g$ and $A_{1g}$ peaks exhibit the higher sheet conductances. Our best MBE film exhibits low $E_g$ and $A_{1g}$ FWHMs of 3.6 and 3.5 $cm^{-1}$, respectively, inducing a record sheet conductance of 1.6 mS for grown films (Figure 4b). The lowest $E_g$ and $A_{1g}$ FWHMs reported are ~ 2.5 $cm^{-1}$ and were obtained with monocrystalline exfoliated $PtSe_2$,[49] however the corresponding transport properties were not studied. This highlights that the sheet conductance of $PtSe_2$ can still be improved and that the intrinsic factors limiting the conductivity remain to be investigated.

In summary, this study successfully demonstrates the relevance and efficiency of using $E_g$ and $A_{1g}$ FWHMs as new Raman metrics for evaluating the $PtSe_2$ quality in terms of both structural and electrical properties. Monitoring the sheet conductance in conjunction with the FWHMs of both $E_g$ and $A_{1g}$ peaks also offer valuable insights into the underlying mechanisms, highlighting their utility for monitoring the synthesis process of $PtSe_2$.

**5. Photodetection and Optoelectronic Mixing at 1.55 µm.** We fabricated DC and RF devices to measure respectively the sheet conductance of $PtSe_2$ channels integrated into devices, and to demonstrate high-frequency photodetectors and OEMs operating at the 1.55 µm telecom wavelength. First, a 14 ML $PtSe_2$ film was synthesized directly on a 2-inch sapphire(0001) substrate by using the following conditions: Tg = 520 °C, Ta = 690 °C and $\Phi$(Se) = 0.3 Å.$s^{-1}$. Note that these conditions are not optimal as the synthesis was not performed under a Se flux of 0.5 Å.$s^{-1}$ (see section 2), however, a film composed of only semimetallic domains is still expected, as demonstrated in section 4. The fabrication process (detailed in Figure S12) includes a first lithographic and etching step to define the different $PtSe_2$ channels and then, a two-step process to contact the channel edges and define the metallic pads. Figure 11a shows the 2-inch sapphire substrate integrating an array of identical cells: measurements were performed on all the devices in the 9 central cells. Each cell contains DC and RF devices.

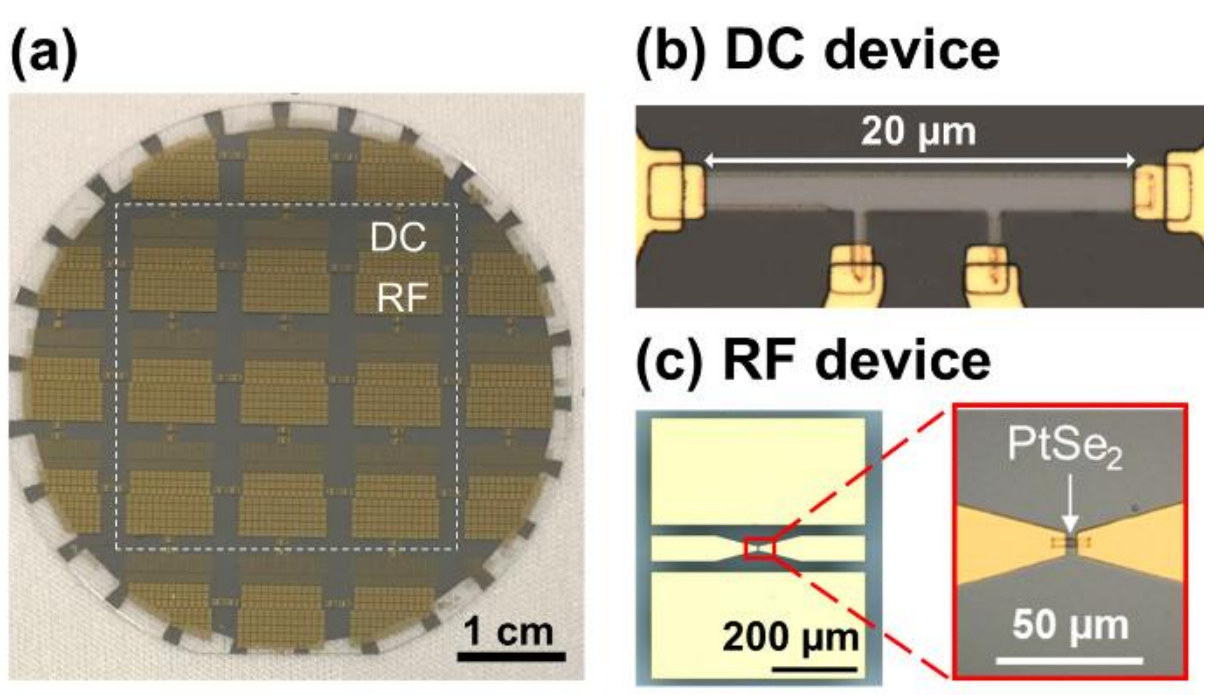

**Figure 11. Optical images of DC and RF devices.** (a) $PtSe_2$-based devices on a 2-inch sapphire substrate. Each cell is identical and integrates DC and RF devices. Measurements were performed on all the devices within the white square. (b) Optical image of a four point contact DC device used to measure the sheet and contact resistance of the $PtSe_2$ channel. (c) Optical image of a 3.5 x 3 µm$^2$ $PtSe_2$ channel integrated into the signal line of a CPW. These devices are used for high-frequency photodetection and mixing.

We first characterized the DC devices (Figure 11b) integrating a 20 x 5.4 µm$^2$ $PtSe_2$ channel (length and width, respectively, are measured at the end of the fabrication process). Using the four point contact measurement method at room temperature on 378 devices from different cells, we obtain a median value of $R_{\square}$ = 0.95 kΩ.□$^{-1}$ (corresponding to $\sigma$ = 1.05 mS) with a standard deviation of 0.20 kΩ.□$^{-1}$ (see statistical measurements and mapping in Figure S13). The measured sheet conductance is close to that obtained on the 20 x 20 mm$^2$ samples, *i.e.* 1.13 - 1.31 mS (for Φ(Se) = 0.3 Å.s$^{-1}$), indicating that the high quality of the $PtSe_2$ film synthesized on a 2-inch substrate is maintained after device fabrication. The contact resistance of the devices obtained by the four point contact method is below the measurement precision (< 100 Ω.µm), such a low value is characteristic of semimetallic $PtSe_2$ films.[38,46]

High-frequency optoelectronic measurements (detailed in Figure S14) were performed at room temperature with coplanar waveguides (CPWs) integrating a 3.5 x 3 $\mu m^2$ $PtSe_2$ channel (RF device in Figure 11c). We first characterized these devices as photodetectors (Figure 12a). As our $PtSe_2$ channel acts as a photoconductor, a DC bias ($V_{DC}$ = 4 V) is applied. The channel is also illuminated by a 1.55 µm laser whose intensity is modulated with a Mach-Zehnder interferometer at frequency $f_{opt}$ varying between 2 and 67 GHz. The photodetected power ($P_{ph}$) is measured as a function of $f_{opt}$ for optical powers ($P_{opt}$) ranging between 14.2 and 41.5 mW. The modulation efficiency of the channel resistance increases with the optical power resulting in an increase in the photodetected signal. Figure 12b shows a flat photodetected signal with a responsivity of around 0.2 $mA.W^{-1}$ (~ 8.5 µA at 41.5 mW) and a 3 dB cutoff frequency of 60 GHz. This 60 GHz bandwidth is a record for TMD-based photodetectors.[80] Ji *et al.*[16] investigated the carrier dynamics in our photoexcited thick $PtSe_2$ films using optical-pump THz probe spectroscopy. An ultrafast photoresponse of 1.4 ps was measured for a pump light at a wavelength of 400 nm, meaning that the cutoff frequency observed in our photodetection measurements may indeed be limited by the photocarrier lifetime.

Devices based on semimetallic photoconductors are not as competitive as conventional 3D semiconductor-based photodiodes in terms of photodetection performance. Due to their semimetallic nature, they present short photocarrier lifetimes and high dark current, which drastically reduces responsivity and signal-to-noise ratio.

In contrast, semimetallic $PtSe_2$ is well suited for high-frequency optoelectronic mixers (OEMs), which mix an input RF signal with a modulated optical power as expressed by the equation in Figure 13a. Such devices require materials with photoconductivity that depends linearly on both the applied bias and the optical power. Semimetallic $PtSe_2$ fulfills these requirements, in addition to providing impedance matching with the RF circuit and a fast relaxation path for photocarriers, allowing for a native ultrafast response. Conversely, state-of-the-art photoconductive OEMs implement degraded GaAs semiconductors, using low temperature growth, to ensure sufficiently short photocarrier lifetimes. Moreover, GaAs-based OEMs have reduced efficiency when operating at the telecom wavelength of 1.55 µm.[81,82]

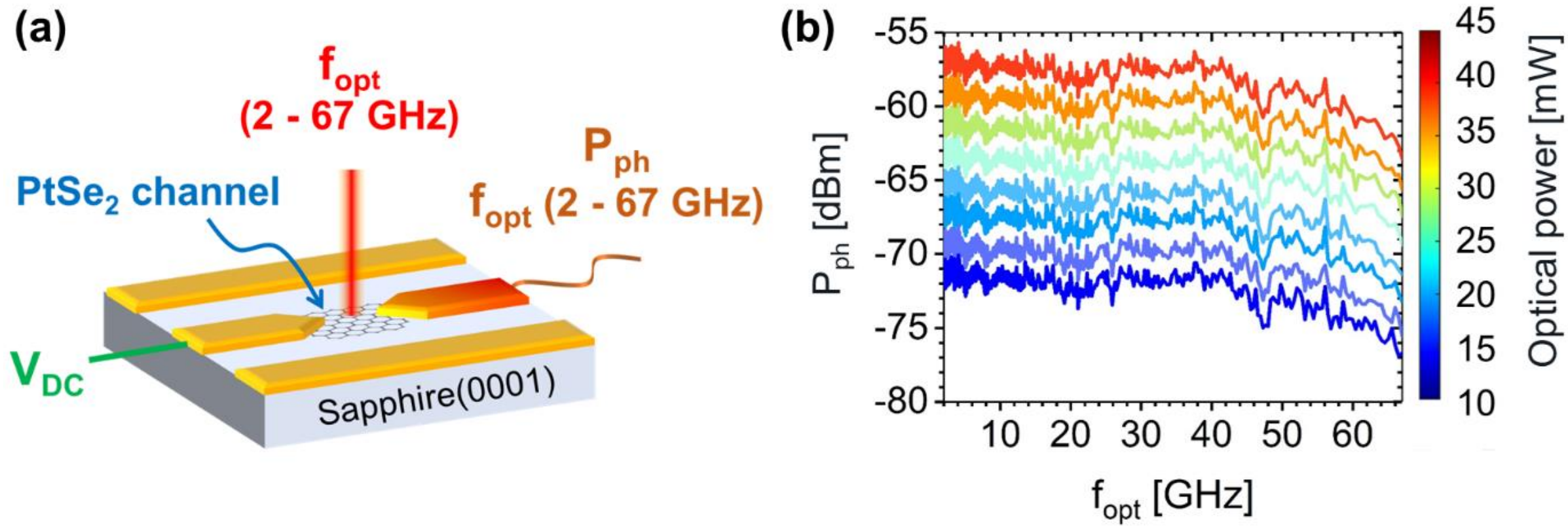

**Figure 12. $PtSe_2$-based 1.55 µm photodetector.** (a) Schematic of the photodetection measurements: CPW integrates a 3.5 x 3 µm$^2$ $PtSe_2$ channel. A DC bias ($V_{DC}$ = 4 V) is applied to the input signal line. A 1.55 µm laser spot, with a diameter of 2.5 µm, is focused on the $PtSe_2$ channel. The laser power is modulated at a frequency ($f_{opt}$) varying between 2 and 67 GHz. (c) Photodetected power ($P_{ph}$) measured as a function of $f_{opt}$ for different optical powers.

Optoelectronic mixing experiments are detailed in Figure S14. When operating as an OEM, the $PtSe_2$ channel bias is provided by the input RF signal. A RF signal with a power $P_{RF}$ = -10 dBm and at frequency $f_{RF}$ varying between 0.01 and 29.99 GHz is injected into the CPW (Figure 13b). The laser power and the modulation frequency are fixed at $P_{opt}$ = 60 mW and $f_{opt}$ = 30 GHz, respectively. The output power ($P_{IF}$) is measured at the intermediate frequency $f_{IF} = f_{opt} - f_{RF}$, which then varies between 29.99 and 0.01 GHz. Figure 13c shows the conversion efficiency, which is the $P_{IF}/ P_{RF}$ ratio. This efficiency is constant over the whole frequency range, which demonstrates a bandwidth above 30 GHz, and reaches -73 dB. This value is 10 dB better than the value obtained by Montanaro *et al.*[46] with CVD graphene CPWs fabricated on a 2-inch $SiO_2$/Si substrate (see Table S3) and highlights the potential of $PtSe_2$ for high-frequency optoelectronics.

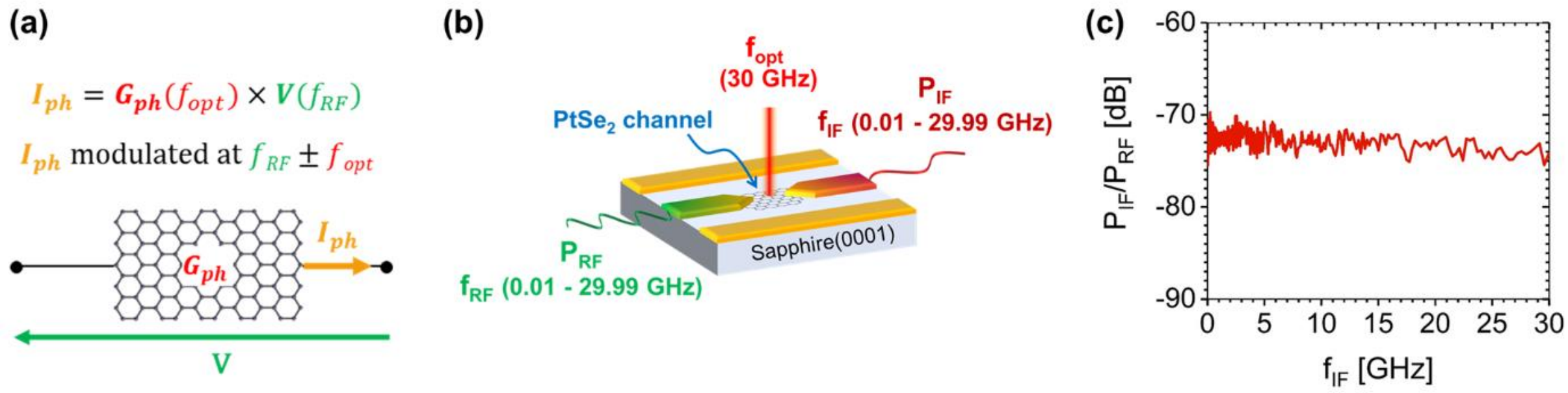

**Figure 13. PtSe2-based 1.55 µm optoelectronic mixer.** (a) Schematic of the optoelectronic mixing mechanism. The output photocurrent ($I_{ph}$) depends both on the voltage (V, modulated at $f_{RF}$) applied on the $PtSe_2$ channel and on the photoconductance ($G_{ph}$, modulated at $f_{opt}$). The output signal includes a signal modulated at $f_{RF} + f_{opt}$ and another one at $f_{RF} - f_{opt}$. (b) Schematic of the optoelectronic mixing experiments. A 1.55 µm laser spot, with a diameter of 2.5 µm, is focused on the $PtSe_2$ channel. The laser power is modulated at a fixed frequency ($f_{opt}$) of 30 GHz. The input RF signal is modulated at a frequency $f_{RF}$ varying between 0.01 and 29.99 GHz. The output power $P_{IF}$ is measured at the intermediate frequency $f_{IF} = f_{opt} - f_{RF}$ varying between 29.99 and 0.01 GHz. (c) Conversion efficiency ($P_{IF}/P_{RF}$) as a function of $f_{IF}$.

## Conclusion

In this study, we quantitatively integrated information from electron microscopy, X-ray diffraction, Raman spectroscopy and electronic transport measurements to optimize the MBE growth of multilayer semimetallic $PtSe_2$ films. We demonstrated that the Raman peak FWHMs of both the in-plane $E_g$ mode and the out-of-plane $A_{1g}$ mode are effective metrics for assessing the structural and electrical properties of multilayer $PtSe_2$ films. Structural analyses revealed that the $E_g$ and $A_{1g}$ FWHMs predominantly reflect the in-plane and out-of-plane crystalline quality, respectively. We showed that the lower the FWHM of both peaks, the higher the crystallinity and the sheet conductance.

Using these Raman metrics in conjunction with the sheet conductance, we optimized the MBE synthesis of thick (12 - 16 MLs) semimetallic $PtSe_2$ films on low-cost and insulating sapphire substrates. A high growth temperature (520 °C), combined with a high Se flux (0.1 - 0.5 $A.s^{-1}$), results in films exhibiting low FWHMs for both the $E_g$ peak ($\leq$ 4 $cm^{-1}$) and $A_{1g}$ peak (< 5 $cm^{-1}$), but limited sheet conductances (0.5 - 0.6 mS). Performing a post-growth annealing at 690 °C under these high Se fluxes yields films with high crystalline quality in both in-plane and out-of-plane directions (exhibiting both $E_g$ and $A_{1g}$ FWHMs < 4 $cm^{-1}$), resulting in record sheet conductances for grown $PtSe_2$ (1.1 - 1.6 mS). The best values of both FWHMs and sheet conductances are obtained at the highest Se flux of 0.5 $A.s^{-1}$, highlighting the importance of a high $\Phi(Se)/\Phi(Pt)$ ratio throughout the synthesis process. Using these optimized growth conditions, we demonstrated a 15 ML MBE film exhibiting $E_g$ and $A_{1g}$ FWHMs down to 3.6 and 3.5 $cm^{-1}$, respectively, along with a high sheet conductance of 1.6 mS, matching the highest reported sheet conductance for a 16 ML exfoliated $PtSe_2$ flake. This demonstrates the effectiveness and relevance of the MBE growth method.

In-depth XRD and STEM analyses revealed the remarkable film's structural evolution induced by the post-growth annealing at high temperature. In-plane defect diffusion and grain boundary migration occur during this annealing, enhancing the local in-plane layer quality, which may be monitored by the slight decrease in $E_g$ FWHM. This in-plane crystal reorganization, in turn, leads to a significant improvement in the out-of-plane crystalline quality across the entire volume of the films, as monitored by the strong reduction in $A_{1g}$ FWHM.

This work also highlights the crucial importance of monitoring the domain thickness in order to control the electrical properties of multilayer $PtSe_2$ films. Our annealed samples (with $E_g$ and $A_{1g}$ FWHMs $\leq$ 4 $cm^{-1}$) are mainly constituted of highly crystalline domains having a thickness equal to the film thickness and exhibit record sheet conductances. In contrast, when the domains are thinner than the overall film (*e.g.* in our as-grown samples with $A_{1g}$ FWHM > 4 $cm^{-1}$), the sheet conductance depends not only on the film thickness but also on the thickness of component domains. Thus, the post-growth annealing at high temperature is essential to achieve a domain thickness equal to the film thickness and to obtain highly conductive $PtSe_2$ films.

Further improvements in the sheet conductance of grown $PtSe_2$ films are expected by enhancing the in-plane crystalline quality, for example by growing epitaxial films to reduce the grain boundary density. This is motivated by the recent achievement of monocrystalline exfoliated $PtSe_2$ exhibiting $E_g$ and $A_{1g}$ FWHMs as low as ~ 2.5 $cm^{-1}$, showing that there is still room for significant improvement in grown $PtSe_2$ crystallinity. Targeting monocrystalline grown $PtSe_2$ is feasible by implementing sapphire vicinal surfaces to promote the epitaxial growth of $PtSe_2$ films; a method already demonstrated for VPT-grown $MoS_2$, which resulted in the in-plane alignment of the domains.[83,84,85]

Finally, we demonstrated optoelectronic devices operating at the telecom wavelength of 1.55 µm, based on our highly crystalline and conductive semimetallic $PtSe_2$ directly synthesized on a 2-inch sapphire wafer. We presented photodetectors with a record bandwidth of 60 GHz for a TMD film and the first TMD-based optoelectronic mixer with a bandwidth above 30 GHz. Considering these landmark results, along with the exceptional air stability observed over 1.5 years with preserved electrical properties, we conclude that $PtSe_2$ is a highly promising material for high-frequency IR optoelectronics.

## Methods

**$PtSe_2$ synthesis.** The growth reactor is a 2-inch MBE system supplied by Dr Eberl MBE-Komponenten. $PtSe_2$ is synthesized on c-plane (0001) sapphire surfaces purchased from Kyocera Corporation (Japan). 20 x 20 $mm^2$ samples are implemented for growth studies (Sections 1 to 4), and a 2-inch substrate was used to demonstrate photodetection and optoelectronic mixing (Section 5). The substrates are first chemically cleaned with acetone/propanol and a 100nm-thick Mo film is deposited on the back to improve the radiative heating of the substrate in the growth chamber. The substrates are outgassed under ultra-high vacuum ($10^{-10}$ mbar) at 500 °C overnight and then, annealed at 900 °C during 15 min. A valved selenium cracker source is used: the tank filled with ultra-high purity (7N) Se is heated to 290 °C and the cracker temperature is maintained at 600 °C. The electron beam evaporator generates a Pt flux from high purity (4N) Pt. Se and Pt fluxes are calibrated prior to synthesis using a quartz crystal microbalance positioned at the sample position. The growth and annealing temperatures listed in the article correspond to the values provided by the thermocouple located behind the sample.

**Raman spectroscopy.** The Raman spectroscope Labram HR Evolution from Horiba is used at room temperature with a 532 nm laser excitation, a laser power of 0.25 mW and a grating of 1800 grooves.mm$^{-1}$. Raman spectra were acquired at 9 positions evenly distributed on each 20 x 20 mm$^2$ sample to extract the median of the $E_g$ and $A_{1g}$ FWHMs plotted in Figure 2 and 3. The plotted vertical lines indicate the first and third quartile of the 9 FWHM values and characterize the inhomogeneity of the $PtSe_2$ films. $A_{1g}$ FWHMs present larger interquartiles compared to $E_g$ FWHMs due to the lower signal-to-noise ratio of the $A_{1g}$ peak.

**EDX measurements.** The ZEISS SIGMA SEM is used with an acceleration voltage of 5 kV, on a scan area of 20 x 20 µm$^2$ at two different positions for each sample.

**XRD measurements.** We implement a Rigaku Smartlab diffractometer equipped with a high brilliance rotating anode and a 5-circle goniometer. In-plane GIXRD measurements are performed without monochromator, with a double tilt wafer stage to maintain a fixed grazing incidence geometry during the sample rotation. The grazing angle is 0.4 °, resulting in an illuminated area of ~ 1 cm$^2$. Incident and diffracted beam parallel slit collimators are used to define a 0.5 ° angular resolution. A Ge(220) 2-bounce monochromator is used for out-of-plane HRXRD measurements. The $PtSe_2$ diffraction peaks are indexed using the file ICDD 03-065-3374, which corresponds to a hexagonal structure with a space group P-3m1 (164).

**STEM imaging.** $PtSe_2$ films are transferred to a TEM grid to perform STEM plane views. First, a 100nm-thick Au layer is evaporated on $PtSe_2$. The Au/$PtSe_2$ stack is peeled off with a standard dicing tape and deposited onto a clean $Si/SiO_2$ wafer. The Au layer is cleaned and then etched with a $KI_2$ solution. The $PtSe_2$ film is picked up with a stamp made of a polypropylene carbonate (PPC) layer spin-coated on a polydimethylsiloxane (PDMS) droplet and released at 120 °C on the TEM grid. It is finally cleaned in acetone/isopropanol to remove polymer residues. STEM analysis are performed using probe aberration corrected microscopes, Titan Themis (Thermofisher) and Titan Ultimate (Thermofisher), operating at 200 kV. The 4D-datasets are obtained using a direct electron detector (Quantum Detector, Medipix-Merlin) equipped in Titan Ultimate, in single pixel continuous mode with a T0 threshold of 30 keV, a convergence angle of 1 mrad and probe size of 1 nm.

**Sheet conductance measurements.** After $PtSe_2$ synthesis, Ti(20 nm)/Au(400 nm) pads of 1.2 mm diameter are evaporated through a shadow mask at each corner of the 18.5 x 18.5 $mm^2$ $PtSe_2$ films grown on the 20 x 20 $mm^2$ substrates. Sheet conductance measurements are then performed using a Hall-effect measurement bench Ecopia HMS 7000 with a current of 1 mA.

## Data Availability

The data presented in this work are available from the corresponding authors E.D. and P.L.

**Acknowledgments**

The authors acknowledge the financial support from the European Union's Horizon 2020 program under grant agreement no. 881603 (Core3 Graphene Flagship), as well as from ANR-18-CE24-0001 (BIRDS) and ANR-20-CE09-0026 (2DonDemand). The authors also acknowledge the financial support from the CNRS-CEA "METSA" French network (FR CNRS 3507) on the platform PFNC (CEA Grenoble) and the French RENATECH network. H. Okuno and D. Dosenovic acknowledge the financial support from the European Union's H2020 Research and Innovation program *via* the e-See project (Grant No. 758385). H. Okuno acknowledges J.L. Rouviere for kindly providing the script used for the GPA analysis, and N. Vigano and A. Camatchy for providing the data processing tools used for structural mapping. E. Desgué and P. Legagneux acknowledge S. Shvarkov from MBE-Komponenten and L. Travers from C2N for their great support and advise to implement the MBE reactor.

**Author Information**

**†Present Addresses**

M.T. – Laboratoire Matériaux et Phénomènes Quantiques, Université Paris Cité, 75013 Paris, France

Z.B. – CRHEA, rue Bernard Grégory, 06560 Valbonne, France

N.M. - Center for Nanoscience and Nanotechnology (C2N), 10 boulevard Thomas Gobert, 91120 Palaiseau, France

## Author contributions

E.D., H.O. and P.L. contributed equally to the manuscript and participated to all the studies. I.V., N.M., Z.B. contributed to the MBE growth. E.B. and M.T. did the optical characterizations. A.O. and J.C. contributed to the Raman characterizations. L.L. did the XRD characterizations. B.M., D.D. and H.O. did the TEM studies. L.G., D.J. and P.P. fabricated the (opto)electronic devices and D.P., D.C. and E.G. measured the devices. B.P. and D.P. contributed to the interpretation of the results.

## Competing Interests

The authors declare no competing interests.

# Supporting Information

**Growth of Highly Conductive $PtSe_2$ Films Controlled by Raman Metrics for High-Frequency Photodetectors and Optoelectronic Mixers at 1.55 µm**

*Eva Desgué[1 *‡], Ivan Verschueren[1†], Marin Tharrault[2†], Djordje Dosenovic[3], Ludovic Largeau[4], Eva Grimaldi[1†], Delphine Pommier[1], Doriane Jussey[1], Bérangère Moreau[3], Dominique Carisetti[1], Laurent Gangloff[1], Patrick Plouhinec[1], Naomie Messudom[1†], Zineb Bouyid[1†], Didier Pribat[5], Julien Chaste[4], Abdelkarim Ouerghi[4], Bernard Plaçais[2], Emmanuel Baudin[2], Hanako Okuno[3‡], Pierre Legagneux[1 *‡]*

## 1. Determination of $PtSe_2$ film thickness by EDX

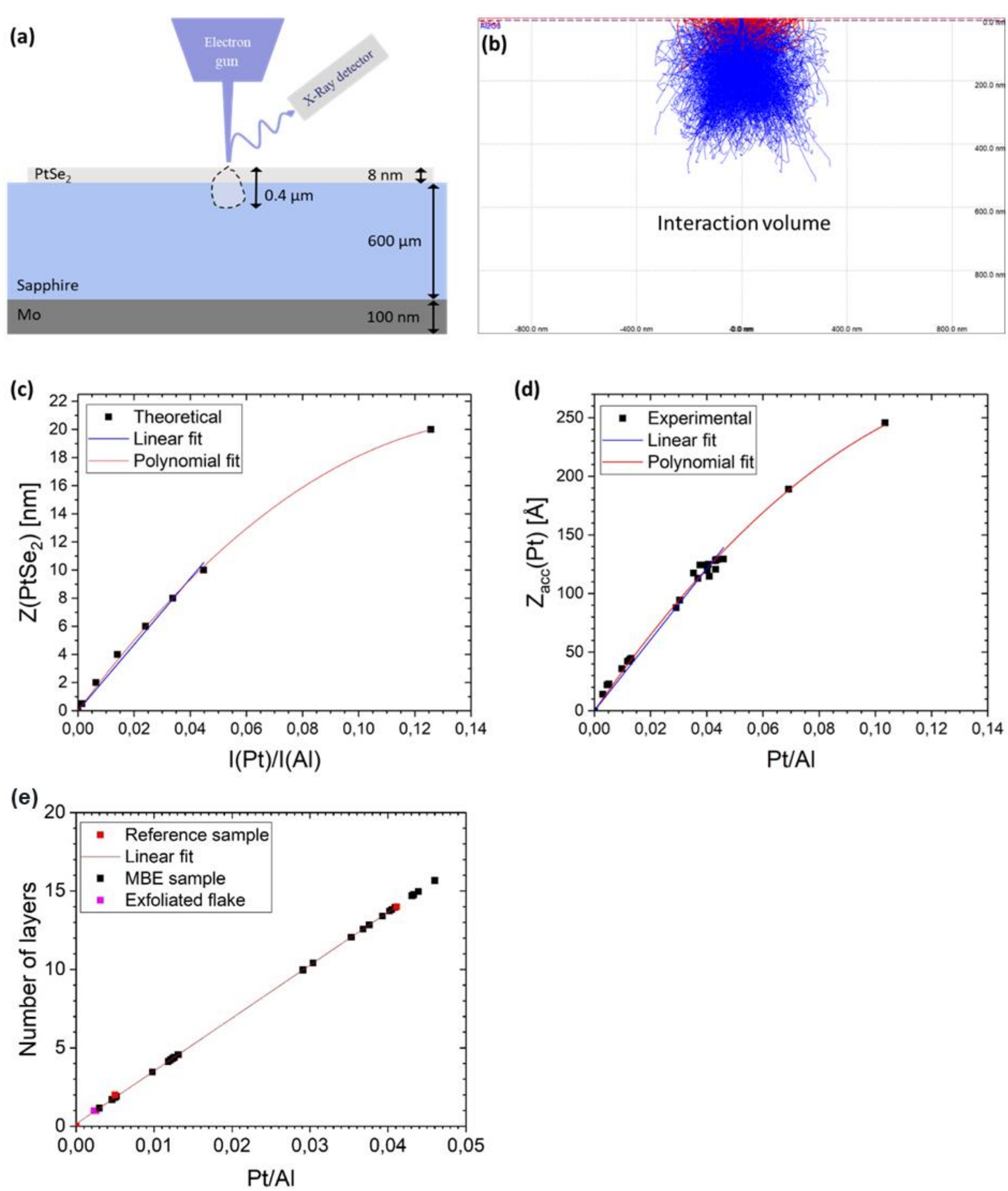

**Figure S1. Determination of $PtSe_2$ film thickness by EDX.** (a) Schematic of the MBE-grown $PtSe_2$/sapphire samples with the electron-matter interaction volume during the EDX characterization. (b) Simulation of the interaction volume resulting from the interaction of 2000 electrons at 5 keV with a $PtSe_2$(8 nm)/sapphire(600 µm) sample, using the CASINO software. The trajectories of backscattered electrons (red) and absorbed electrons (blue) are shown. (c)

$PtSe_2$ film thickness $Z(PtSe_2)$ as a function of the simulated intensity ratio between the Pt (M) X-ray and the Al (K) X-ray: I(Pt)/I(Al). All the data are fitted with a second-degree polynomial function (red curve), whereas the data for $Z(PtSe_2)$ ranging from 0 to 9 nm are fitted with a linear function (blue curve). (d) Experimental values of the Pt thickness accumulated on the QCM during the $PtSe_2$ growth step ($Z_{acc}(Pt)$) as a function of the Pt/Al EDX atomic ratios of the corresponding $PtSe_2$/sapphire samples. All the data are fitted with a second-degree polynomial function (red curve), whereas the data for Pt/Al ranging from 0 to 0.05 are fitted with a linear function (blue curve). (e) Number of layers of MBE-grown $PtSe_2$/sapphire samples as a function of their Pt/Al EDX atomic ratios. The number of layers of two reference samples (red squares) were extracted by optical absorption measurements (see method in Tharrault *et al.*[1]) and the results were fitted with a linear function (red line). The number of layers of our MBE samples (black squares) was calculated from the equation of the red line and their Pt/Al ratios. The number of layers of the exfoliated flake (purple squares) was determined by Raman spectroscopy.[2]

To determine the film thickness ($Z(PtSe_2)$) of our $PtSe_2$ films grown on sapphire ($Al_2O_3$) substrates, we have used the atomic ratio of Pt/Al measured by EDX: Pt comes from the $PtSe_2$ film and Al comes from the substrate. This ratio is related to the film thickness as demonstrated by the simulations in Figure S1. We simulated the corresponding electron-matter interaction with $Z(PtSe_2)$ ranging from 0.5 to 20 nm, using the software CASINO. First, at $Z(PtSe_2)$ = 8 nm (which is our main film thickness), the electron-matter interaction depth is ~ 400 nm, meaning that the electron-$PtSe_2$ interaction depth is only 1/50 of the total interaction depth, as shown in Figure S1(a-b). EDX characterization uses the intensity of the non-absorbed Pt (M) X-rays and Al (K) X-rays (I(Pt) and I(Al), respectively) to calculate the atomic percentages of Pt and Al present in the interaction area. This means that I(Pt)/I(Al) is directly related to Pt/Al. Therefore, we have secondly simulated I(Pt) and I(Al) for $Z(PtSe_2)$ ranging from 0.5 to 20 nm to calculate the corresponding I(Pt)/I(Al) theoretical ratios, as shown in Figure S1c (black dots). These data are well fitted by a second-degree polynomial function (red curve). Moreover, for $Z(PtSe_2)$ ranging from 0 to 9 nm, which is the film thickness investigated in the present study, I(Pt)/I(Al) is well fitted by a linear function (blue curve). Therefore, the Monte Carlo simulations demonstrate that the Pt/Al atomic ratio measured by EDX is linearly correlated to $Z(PtSe_2)$ when $Z(PtSe_2)$ is $\leq$ 9 nm.

Experimentally, we plotted the Pt thickness accumulated on the quartz crystal microbalance (QCM) during the $PtSe_2$ growth step ($Z_{acc}$(Pt)) of different MBE samples, which is directly related to Z($PtSe_2$), as a function of their EDX atomic ratio of Pt/Al (Figure S1d). The data are well fitted by a second-degree polynomial function (red curve), confirming that the Pt/Al ratio is a good metric to assess Z($PtSe_2$). It also shows that Pt/Al is linearly linked to $Z_{acc}$(Pt), and thus Z($PtSe_2$), for Pt/Al ranging from 0 to ~ 0.05 (blue curve), which is in good agreement with the simulations (Figure S1c).

Note that during the EDX characterization, the scanned areas of the $PtSe_2$ samples are 20 x 20 $\mu m^2$, meaning that the measured atomic percentage of Pt takes into account the variations in film thickness over this area. As shown in Figure 9 of the main text, the film thickness of $PtSe_2$ samples can be inhomogeneous over small lateral distances (50 nm). Therefore, the measured Pt/Al EDX ratio is averaged over 20 x 20 $\mu m^2$. To calculate the number of layers (Z($PtSe_2$)/0.51nm) of our MBE-grown samples with their Pt/Al ratios, we used reference samples. The number of layers of two $PtSe_2$ samples was extracted by optical absorption measurements (see method in Tharrault *et al.*[**Erreur ! Signet non défini.**]). The beam spot is ~ 1 $\mu m^2$, meaning that the number of layers extracted by this method also takes into account the variation in film thickness over this area. The number of layers of the two reference samples was determined to be 2 and 14 MLs. These values are plotted in Figure S1e as a function of the corresponding Pt/Al ratios (red squares) and are fitted with a linear function (red line). The extracted equation of the line is used to calculate the number of layers of our MBE $PtSe_2$ samples using their Pt/Al ratios, for Pt/Al values $< 0.05$ (black squares). To show the relevance of this calibration curve, we have also plotted the Pt/Al ratios of a monolayer monocrystalline flake obtained by mechanical exfoliation (ME) (purple squares), which fit very well with the calibration curve.

## 2. Optimization of $PtSe_2$ synthesis

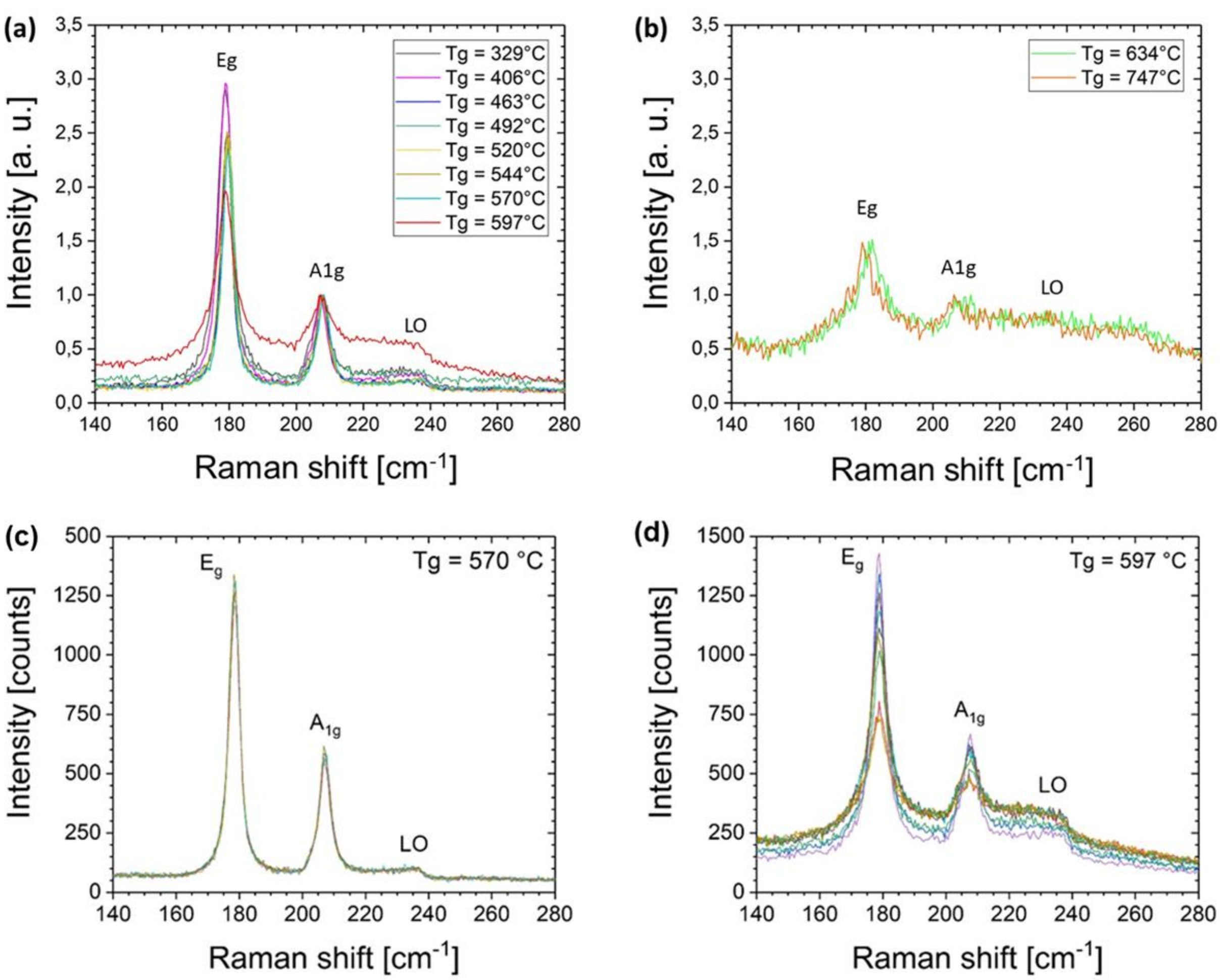

**Figure S2. Raman spectra of as-grown samples synthesized at different growth temperatures.** (a - b) Raman spectra ($\lambda = 532$ nm) of $PtSe_2$ films synthesized at different growth temperatures (Tg) under 0.2 $Å.s^{-1}$ of Se. The intensity of the spectra is normalized to the $A_{1g}$ peak. The Raman shift of the different spectra is not calibrated, the peak position is thus arbitrary. (c - d) Raman spectra acquired at 9 different positions (evenly distributed across the 20 x 20 mm² samples) of the $PtSe_2$ films synthesized at Tg = 570 °C and Tg = 597 °C, respectively.

Figure S2(a-b) show that the 3 characteristic Raman peaks of $PtSe_2$ are observed for all the Tg values. However, from Tg = 597 °C (red curve in Figure S2a), the background level and the FWHM of the $E_g$ and $A_{1g}$ peaks increase significantly, and from Tg = 634 °C (green curve in Figure S2b), the Raman signal becomes very weak and the Raman peaks are no longer well defined. This indicates that the crystalline quality deteriorates from Tg = 597 °C. Moreover,

Figure S2d shows that the Raman spectra acquired at 9 different positions of the sample synthesized at Tg = 597 °C are highly inhomogeneous (for comparison, Figure S2c displays the 9 Raman spectra of the sample synthesized at Tg = 570 °C). This inhomogeneity is highlighted by the strong dispersion of both the $E_g$ and $A_{1g}$ FWHM values (large vertical bars in Figure 2b of the main text) and indicates a significant film inhomogeneity.

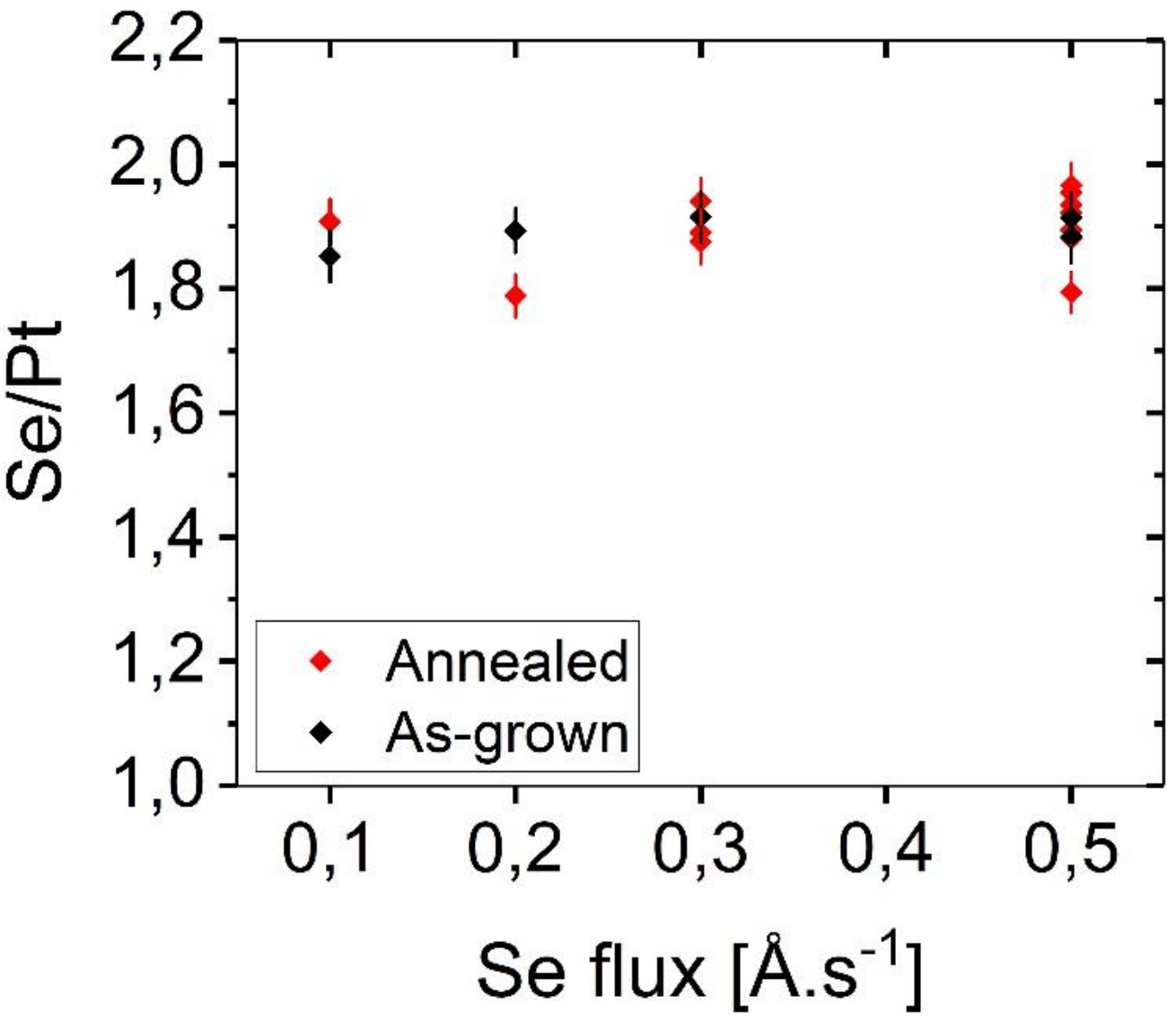

**Figure S3. EDX analysis of as-grown and annealed samples.** Se/Pt atomic ratios of MBE-grown $PtSe_2$ films extracted from EDX measurements. Samples are synthesized at 520 °C without (black dots) or with (red dots) a post-growth annealing at 690 °C for 30 min, under different Se fluxes. The vertical bars correspond to the errors on the Se/Pt ratio values.

Figure S3 shows that there is no obvious relationship between the Se/Pt EDX atomic ratio of samples and the Se flux used during the synthesis. There is also no significant change in the ratio between as-grown and annealed samples, demonstrating that no desorption of Se occurs during the post-growth annealing under Se flux.

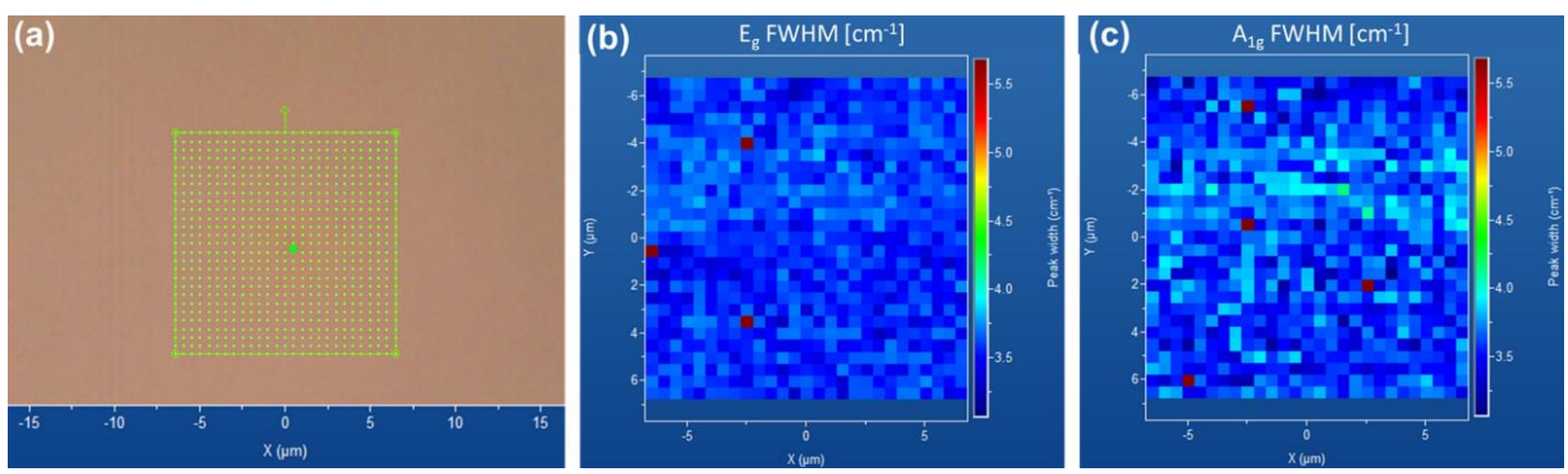

**Figure S4. Raman mapping ($\lambda$ = 532 nm) of an annealed $PtSe_2$ film grown under 0.5 Å.$s^{-1}$ of Se.** (a) is the optical image of the analyzed area (13 x 13 $\mu m^2$). (b) and (c) show the corresponding mapping of the $E_g$ and $A_{1g}$ FWHMs, respectively. The red pixels are not included in the statistical analysis and are due to a spurious Raman peak that distorts the $E_g$ or $A_{1g}$ peak.

## 3. Air stability of $PtSe_2$

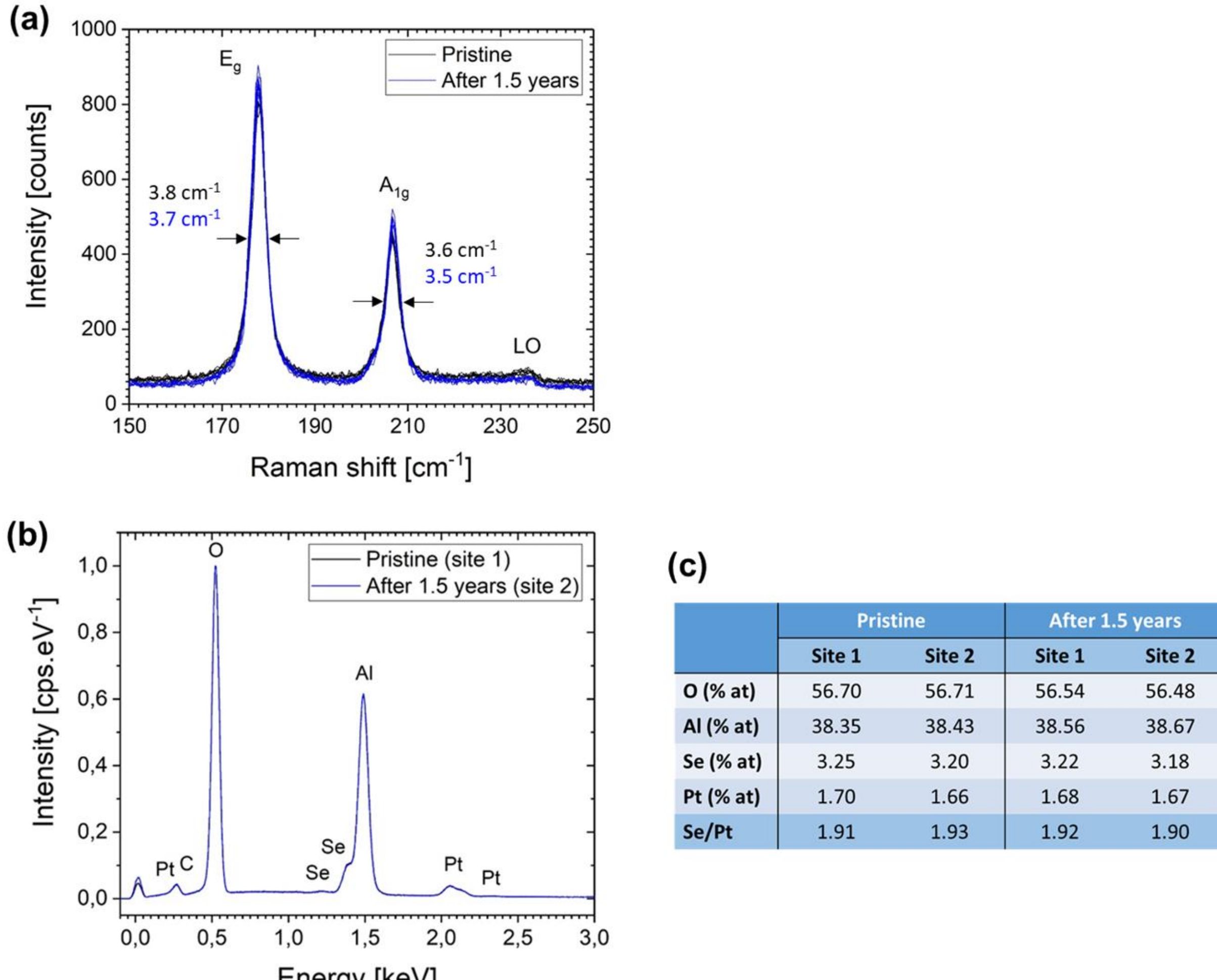

| | Pristine | | After 1.5 years | |
|---|---|---|---|---|
| | Site 1 | Site 2 | Site 1 | Site 2 |
| O (% at) | 56.70 | 56.71 | 56.54 | 56.48 |
| Al (% at) | 38.35 | 38.43 | 38.56 | 38.67 |
| Se (% at) | 3.25 | 3.20 | 3.22 | 3.18 |
| Pt (% at) | 1.70 | 1.66 | 1.68 | 1.67 |
| Se/Pt | 1.91 | 1.93 | 1.92 | 1.90 |

**Figure S5. Air stability of $PtSe_2$ demonstrated by Raman and EDX analysis.** (a) Raman spectra ($\lambda$ = 532 nm) recorded on an MBE-grown annealed $PtSe_2$ sample a few days after the synthesis (black curves) and after 1.5 years in air (blue curves). The corresponding median FWHM values of the $E_g$ and $A_{1g}$ peaks are displayed. (b) EDX spectra of the corresponding $PtSe_2/Al_2O_3$ sample measured a few days after the synthesis (black curve) and after 1.5 years in air (blue curve). (c) Table showing the atomic percentages extracted from the EDX spectra and the calculated Se/Pt atomic ratios from two different analyzed areas (site 1 and site 2) of the sample.

## 4. Structural analysis of the as-grown and annealed samples

a. GIXRD φ scan

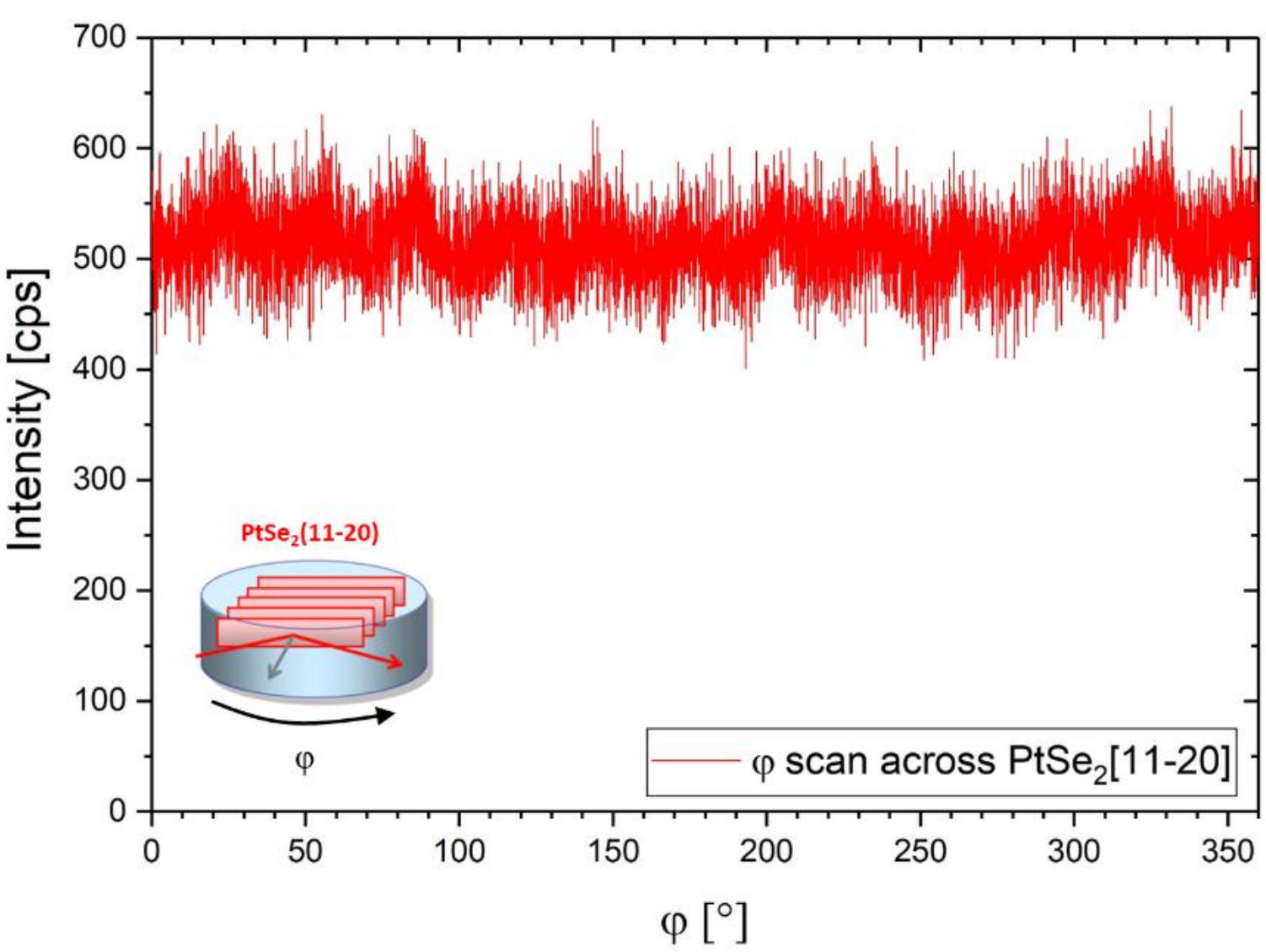

**Figure S6. GIXRD azimuthal (φ) scan across the $PtSe_2$[11-20] direction of an annealed $PtSe_2$ film.** The X-ray detector was positioned at the angle corresponding to the diffraction of the $PtSe_2$(11-20) planes and the sample was rotated azimuthally over 360 °. The intensity of the $PtSe_2$(11-20) diffraction peak is plotted as a function of the azimuthal angle (φ). The isotropic signal indicates that the $PtSe_2$ domains have random in-plane orientations.

## b. Domain size estimation by GIXRD

The in-plane domain size of the samples was first calculated using the Scherrer formula (S1), which resulted in size values (D) exhibiting a strong variation with the diffraction angle,[3] as shown in Table S1. The Scherrer formula neglects the contribution of non-uniform lattice distortions (microstrains), which also produce significant broadening of the X-ray diffraction peaks.[4]

**As-grown sample**

| hkil | $2\theta_\chi$ [°] | $\Delta(2\theta_\chi)_{meas}$ [°] | **D [Å]** |
|---|---|---|---|
| 10-10 | 27.82031 | 0.5252 | 255.7617409 |
| 11-20 | 49.23445 | 0.6579 | 176.1212809 |
| 20-20 | 57.53772 | 0.7215 | 158.6028808 |
| 30-30 | 92.42317 | 1.1502 | 111.4334124 |
| 22-40 | 112.822 | 1.5485 | 100.3962598 |

**Annealed sample**

| hkil | $2\theta_\chi$ [°] | $\Delta(2\theta_\chi)_{meas}$ [°] | **D [Å]** |
|---|---|---|---|
| 10-10 | 27.70473 | 0.481 | 329.4913909 |
| 11-20 | 48.99778 | 0.548 | 247.1201257 |
| 20-20 | 57.17303 | 0.5964 | 215.9398321 |
| 30-30 | 91.81508 | 0.8539 | 158.702245 |
| 22-40 | 112.025 | 1.0445 | 154.2406319 |

**Table S1. Analysis of the in-plane domain size of as-grown and annealed $PtSe_2$ films.** The in-plane domain size (D) of the samples is calculated using the Scherrer formula:

$$D = \frac{K\lambda}{\Delta(2\theta_\chi)\cos(\theta_\chi)} \quad \text{(S1)}$$

where K is the Scherrer constant (0.94 for isotropic crystallites), $\lambda$ is the X-ray wavelength (1.54056 Å), $2\theta_\chi$ is the diffraction angle and $\Delta(2\theta_\chi)$ is the peak FWHM deconvoluted from the instrument width, such that:

$$\Delta(2\theta_\chi) = \sqrt{\Delta(2\theta_\chi)^2_{meas} - \Delta(2\theta_\chi)^2_{instru}} \quad \text{(S2)} \quad \text{with } \Delta(2\theta_\chi)_{instru} = 0.405\,°$$

Equation S1 gives domain size values (D) exhibiting a strong variation with the diffraction angle ($2\theta_\chi$) for both samples. For example, the FWHM ($\Delta(2\theta_\chi)_{meas}$) of the $PtSe_2$(11-20) and $PtSe_2$(22-40) diffraction peaks of the annealed sample give calculated D values of 25 and 15 nm, respectively, even though these planes belong to the same family. Such a contradictory result reveals that not only the limited size of the crystallites is responsible for the diffraction peak broadening. Therefore, the classical Scherrer formula cannot be reliably applied in this case.

We then implemented the Williamson-Hall method[5] to evaluate the contribution of the in-plane domain size ($D_{WH}$) and the lattice microstrain ($\varepsilon$) to the peak broadening (see e.g. for $WSe_2$[6]). According to this model:

$$\Delta(2\theta_\chi)_{size} = \frac{K\lambda}{D_{WH}\cos(\theta_\chi)} \quad (S3)$$

$$\Delta(2\theta_\chi)_{strain} = 4\mathcal{E}\tan(\theta_\chi) \quad (S4)$$

Thus, injecting equation S3 and S4 into equation S1 gives:

$$\Delta(2\theta_\chi)\cos(\theta_\chi) = \frac{K\lambda}{D_{WH}} + 4\mathcal{E}\sin(\theta_\chi) \quad (S5)$$

In Figure 5b of the main text, $\Delta(2\theta_\chi)\cos(\theta_\chi)$ is plotted as a function of $\sin(\theta_\chi)$ and fitted with a linear function. For both the as-grown and annealed samples, the experimental data are well fitted by linear functions, as predicted by the method. From the slope and y-intercept, we derive $D_{WH}$ and Ɛ values which are shown in Table S2.

| | Intercept = $K\lambda/D_{WH}$ | Slope = 4Ɛ | $D_{WH}$ [Å] | Ɛ [%] |
|---|---|---|---|---|
| As-grown | 0.002 ± 1.477E-4 | 0.01501 ± 2.54957E-4 | 724 ± 54 | 0.38 ± 0.0064 |
| Annealed | 0.0023 ± 3.89448E-4 | 0.00895 ± 6.75578E-4 | 630 ± 107 | 0.22 ± 0.017 |

**Table S2. Williamson-Hall plot data for the as-grown and annealed samples.** $D_{WH}$ is the in-plane domain size and Ɛ the lattice microstrain.

c. Grain boundary structures in the annealed sample

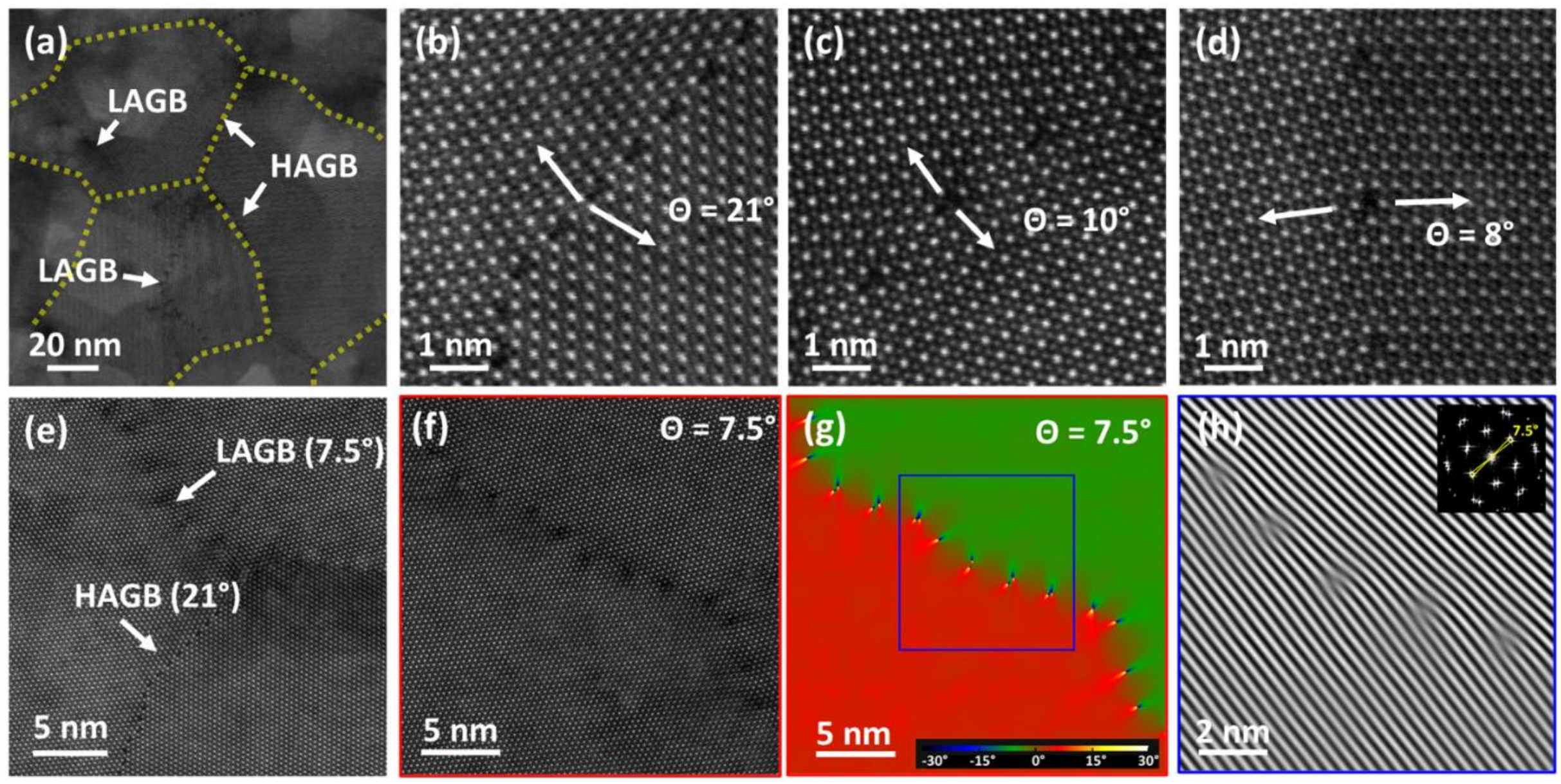

**Figure S7. High resolution STEM images of grain boundary structures.** (a - f) HAADF-STEM images of grain boundaries between grains misoriented by an angle θ, observed in the annealed $PtSe_2$ sample. (g) Orientation field map obtained by Geometric Phase Analysis (GPA) for the area shown in (f). The color scale corresponds to the relative in-plane angle. (h) Dislocation cores visualized by FFT filtering for the $PtSe_2$(10-10) spot, as shown in the inset.

As shown in Figure S7a and S7e, two types of grain boundaries (GBs) are observed in the annealed sample. The first type consists of crystalline domains with in-plane sizes ranging from 50 to 80 nm, surrounded by continuous line defects, as indicated by the yellow dashed lines. These line defects are generally GBs formed between two crystalline grains misoriented by relatively high angles ($\theta > \sim 10$ °), the so-called HA (high angle) GBs. HAGBs are typically composed of a combination of 5 - 7 rings, as studied in the literature for monolayer $PtSe_2$.[7] In these crystalline domains, small-angle misorientations are additionally investigated. The neighboring domains misoriented by relatively small angles ($\theta < \sim 10$ °) are connected by LA (low angle) GBs consisting of several dislocation cores aligned in a line, as shown in Figure S7c, S7d and S7f. The standoff distance between dislocation cores increases as the misorientation angle decreases. The crystal structures are not completely disrupted, as the crystal planes remain connected across the two domains, even in the presence of dislocations, (as shown by GPA analyses in Figure S7(g-h)).

d. HR-STEM dark field imaging

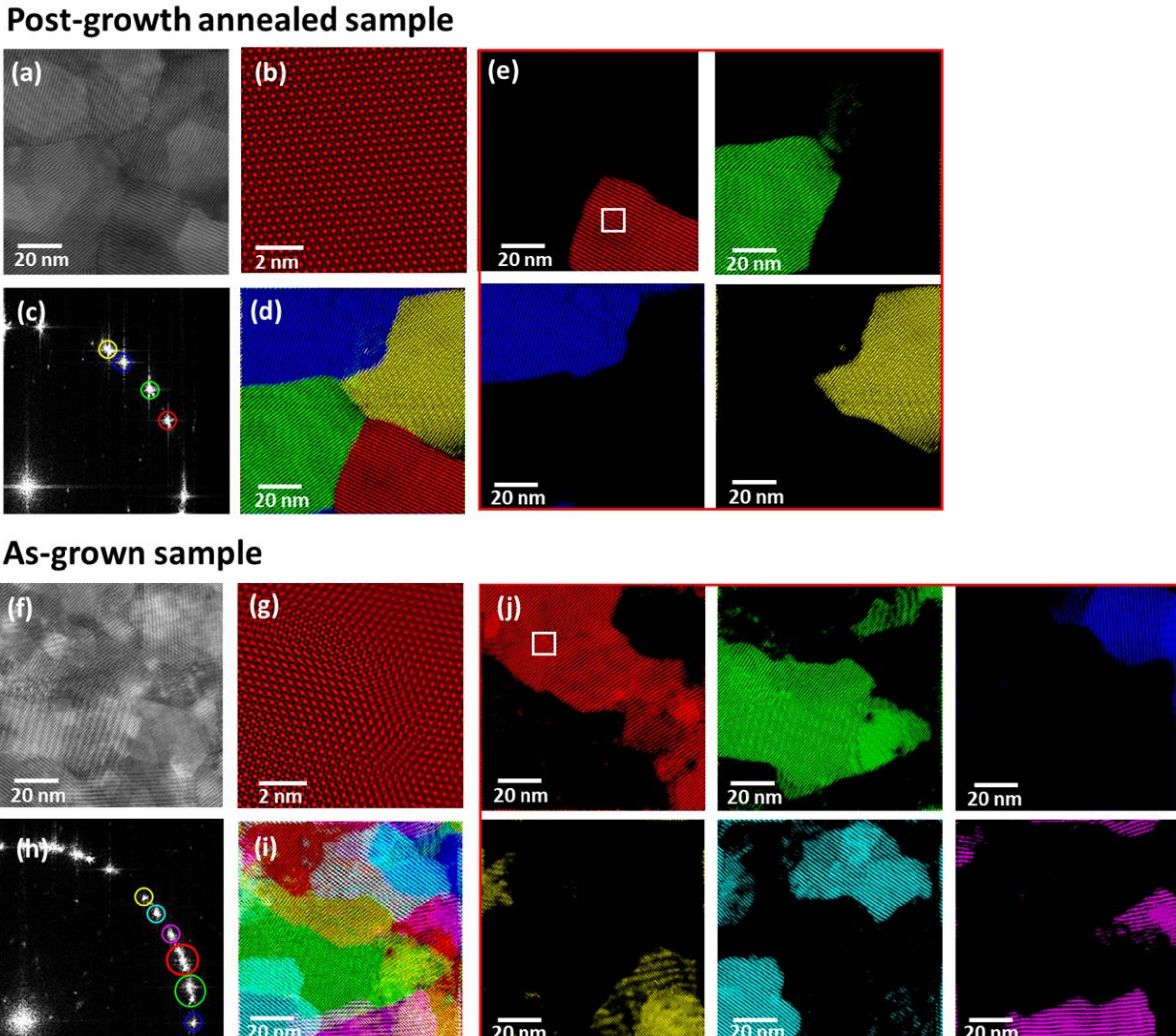

**Figure S8. High resolution STEM dark field images.** (a) and (f) are HAADF-STEM images of the annealed and as-grown $PtSe_2$ samples, respectively. (d) and (i) are the corresponding superimposed DF images obtained from selected FFT spots shown in (c) and (h), respectively. (b) and (g) show typical projected structures in an individual domain, extracted from the white squares in the domain indicated in red in (e) and (j), respectively. (e) and (j) are individual DF images for a selected area in the FFT indicated by the color codes in (c) and (h), respectively.

Figure S8 shows atomic resolution images of the annealed and as-grown samples and the corresponding FFT-based DF images. The annealed sample consists of a lateral conjunction of domains clearly identified by a single spot in the FFT (Figure S8(c-e)), with sharp GBs. In contrast, in the as-grown sample, several domains are identified where 2 to 4 domains with different in-plane orientation ranges are locally superimposed within the film thickness, as shown in Figure S8i. These domains are often not perfectly single crystalline, but contain slight local in-plane misorientation or defects in individual layers. For example, two domains illuminated in red and green in Figure S8j exhibit several spots spread within a small angular range (7.5 °), as indicated by red and green circles in the FFT (Figure S8h). Figure S8g also shows a sliding effect in the projected atomic images which suggests defective feature of as-grown domains.

e. 4D-STEM dark field imaging

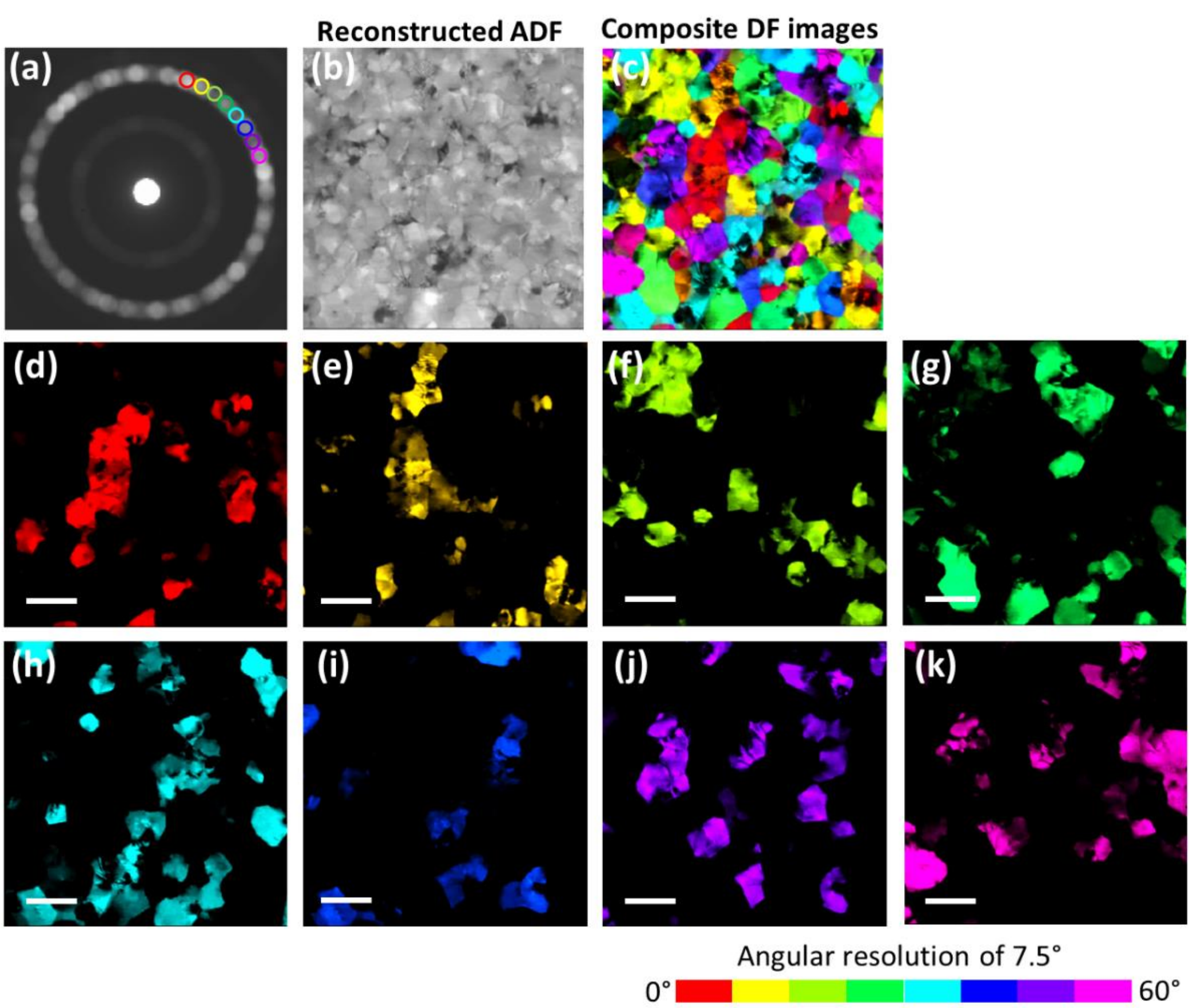

**Figure S9. 4D-STEM DF analysis of the post-growth annealed sample shown in Figure 8 of the main text.** (a) Total diffraction pattern and (b) virtual ADF image. (c) Superimposed and (d - k) separated DF images illuminated for the corresponding angular range (7.5 °) in the diffraction pattern. The color code is shown in the color bar.

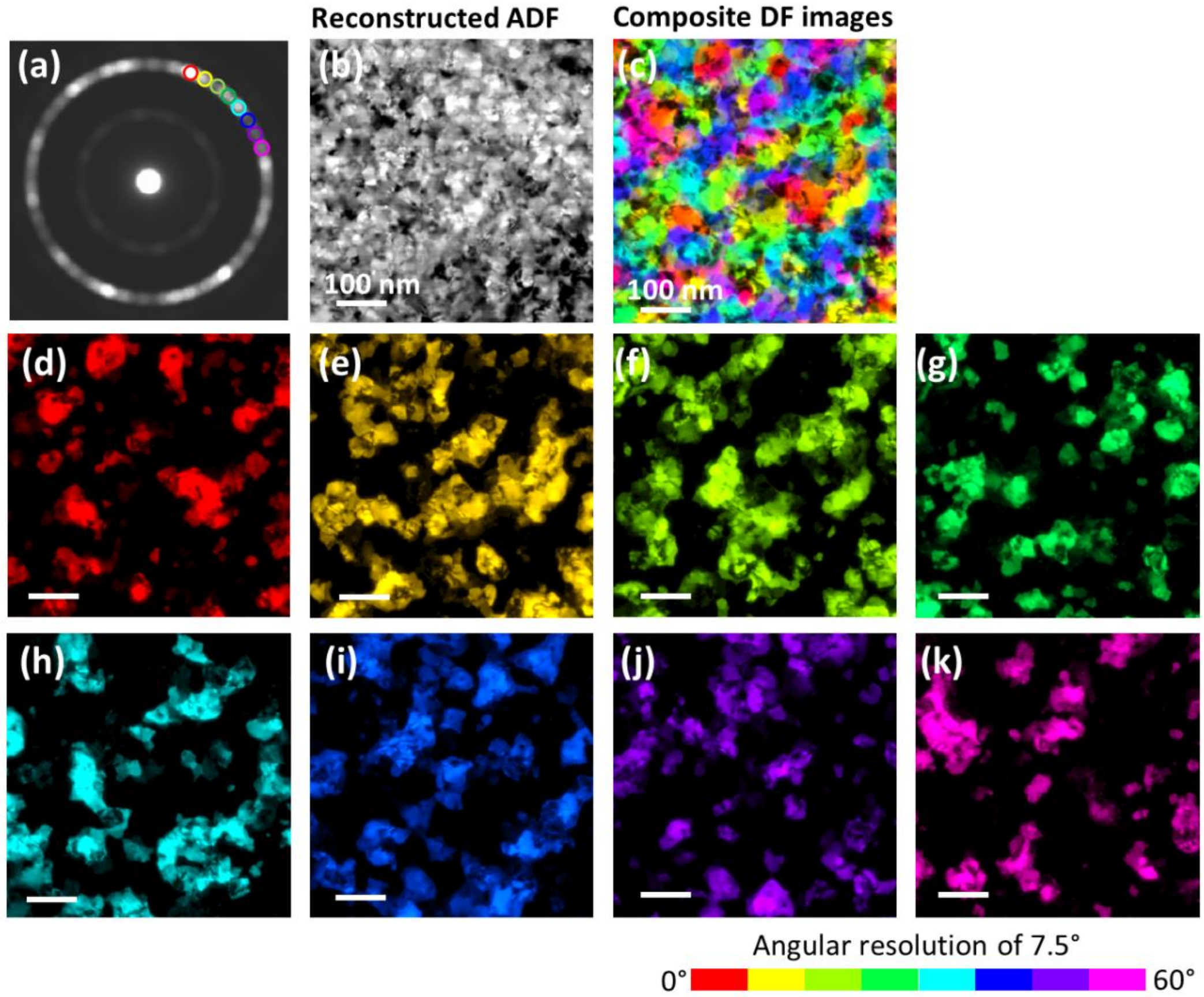

**Figure S10. 4D-STEM DF imaging of the as-grown sample shown in Figure 8 of the main text.** (a) Total diffraction pattern and (b) virtual ADF image. (c) Superimposed and (d - k) separated DF images illuminated for the corresponding angular range (7.5 °) in the diffraction pattern. The color code is shown in the color bar.

The 4D-datasets used for the orientation mapping in Figure 8d and 8h of the main text were also analyzed using filtering. Figure S10 and S11 show DF images of the annealed and as-grown samples, respectively, obtained by the signal separation for a selected angular range (7.5 °). Figure S10 shows that the annealed film is relatively well separated into angular selected domains in the in-plane direction. On the contrary, Figure S11 shows that, in the as-grown sample, several domains illuminated with different angular ranges are often found at the same position. This indicates that the domains with different in-plane orientations are distributed in both lateral and vertical directions. The domain size determined with an angular range of 7.5 ° was between 50 and 80 nm. It should be noted that within 7.5 ° the high-resolution analysis shows a widely spread angular variation (Figure S8h). This confirms that the domains are not really single crystalline, but should contain an important amount of defects, such as dislocations and LAGBs, which could lead to local in-plane strain, as shown by the GIXRD analysis. This large area analysis confirms: i) the superimposition of twisted domains in the as-grown sample and ii) the improvement in the vertical alignment of domains induced by the annealing process.

f. Determination of characteristic standard deviation of the tilt($\vec{c}$)

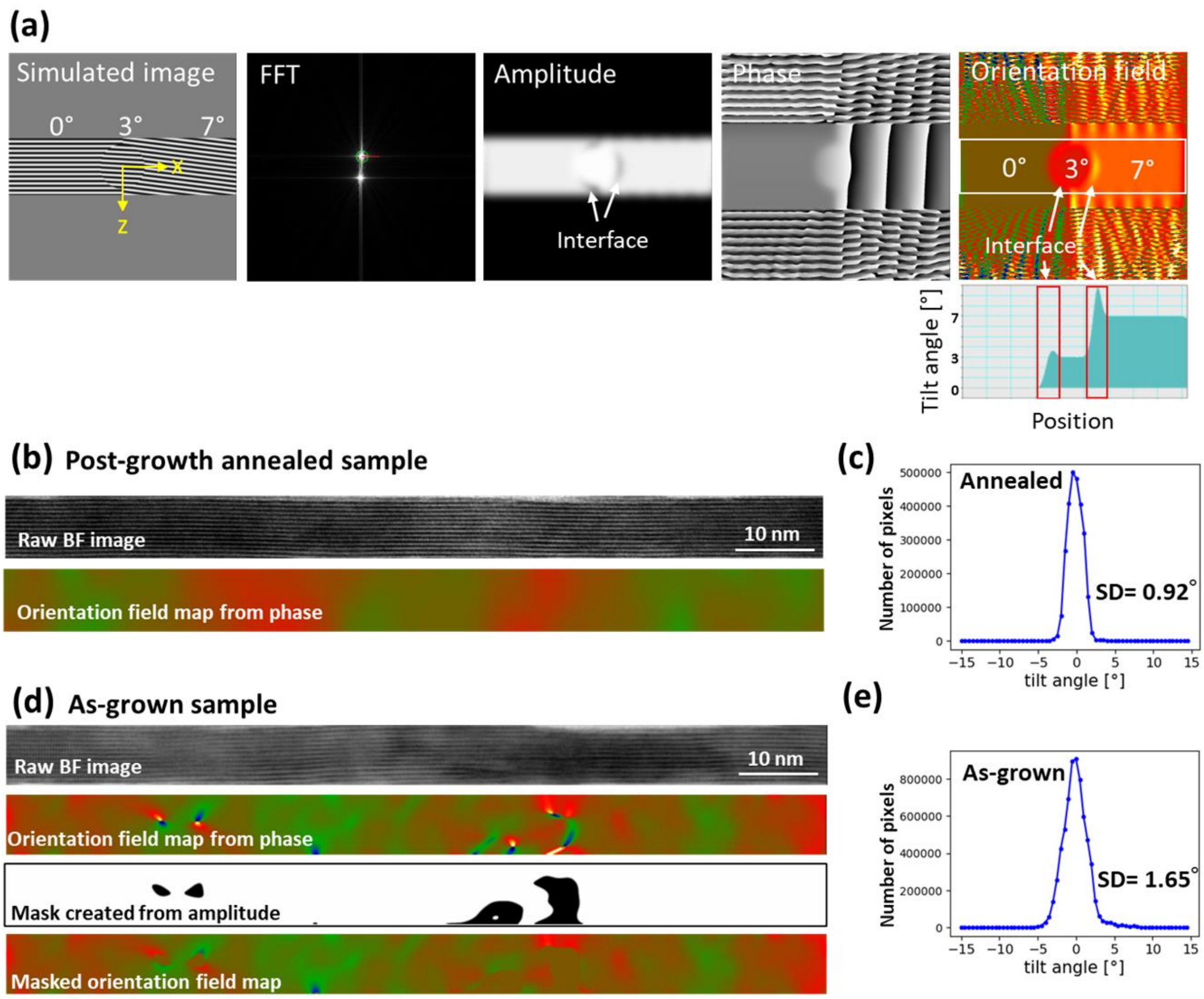

**Figure S11. Determination of the SD tilt($\vec{c}$) value.** (a) GPA analysis process using a simulated interface model. (b) Bright field (BF) image and corresponding orientation field map of the annealed sample and (c) histogram of tilt($\vec{c}$) values extracted from the orientation field map with the characteristic standard deviation (SD) value. (d) BF image, corresponding orientation field map of the as-grown sample, mask created to remove interface effects and the resulting masked orientation field map. (e) Histogram of tilt($\vec{c}$) values extracted from the masked orientation field map with the characteristic standard deviation (SD) value.

To quantitatively assess the flatness of the films, the standard deviation (SD) of the local out-of-plane tilt angle with respect to the sapphire c-plane, tilt($\vec{c}$), was measured from cross-sectional STEM images using GPA.[8,9] Bright field (BF) STEM images were used for better image contrast. The orientation field map, consisting of quantitative values of tilt($\vec{c}$) at each pixel, was generated using the GPA plug-in (digital micrograph) and these values are then collected to create histograms and further calculate the SD tilt($\vec{c}$). Although the local tilt($\vec{c}$) values obtained from the GPA orientation field map are reliable, the artifact that appears at the interface between different angles produces artificially intense lines. Figure S12a shows the process flow for obtaining the orientation field maps, where the simulated structural model, consisting of 3 domains with different angles, is used to examine the interface effect. As shown in the line profile extracted from the orientation field map, the artificially intense lines appear at the interface (indicated by the red squares). To remove these artifacts from the statistical local tilt($\vec{c}$) analysis, we created a mask based on the amplitude image using the threshold optimized with simulated images. Figure S12b and S12d show the orientation field maps, as well as the original BF STEM images, of the annealed and the as-grown films respectively. The annealed film generally shows unidirectional (in-plane direction) corrugation, while the as-grown film shows a more frequent variation of the tilt($\vec{c}$) in both in-plane and out-of-plane directions. This often induces artificial intense lines due to the interface effect. The mask is then used for the as-grown sample to remove these artifacts, in order to collect only the tilt($\vec{c}$) values that originate from the local layer orientation. Figure S12c and S12e display the histograms of tilt($\vec{c}$) values collected at each pixel in several images, covering a total analyzed region in length of 600 nm to ensure reliable statistical analysis, where the collection area length (600 nm) was chosen for a reliable comparison with SD tilt($\vec{c}$) values for TAC samples given in the literature[10] (presented in section 2.4 of the main text). The characteristic SD tilt($\vec{c}$) values over 600 nm obtained from these histograms are 0.92 ° and 1.65 ° for the annealed and the as-grown films, respectively. The number of pixels in the masked areas was removed from the total population for these calculations. It should be noted that reducing the collection area length to around the in-plane domain size allows the SD tilt($\vec{c}$) values to better reflect local out-of-plane tilt disorders, such as the planarity of individual layers comparable to the "c" values measured by HRXRD.

## 5. Device fabrication and measurements

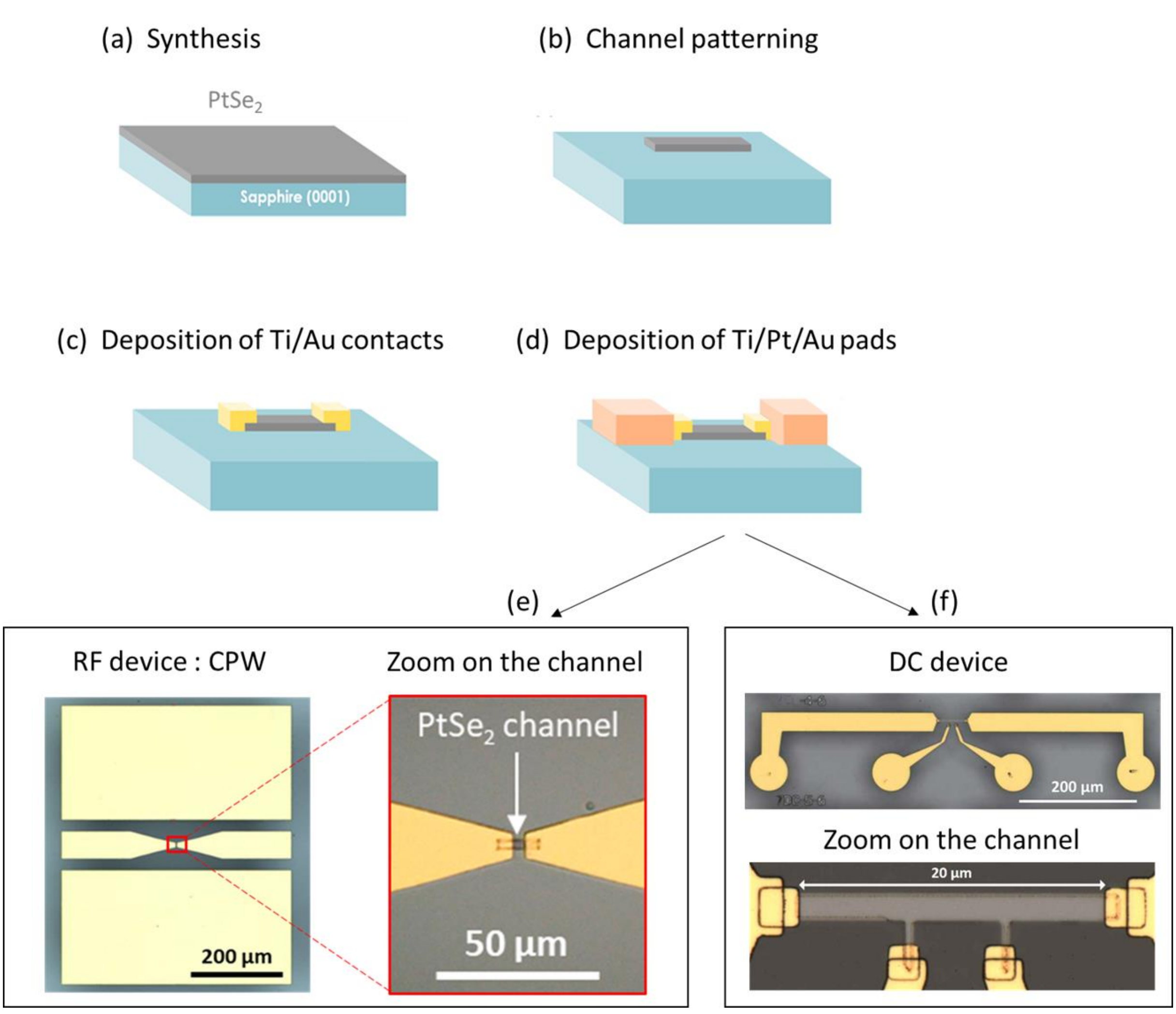

**Figure S12. Fabrication steps of the $PtSe_2$-based devices.** (a) A 14 ML $PtSe_2$ film was synthesized by MBE on a 2-inch sapphire substrate, (b) $PtSe_2$ channels were patterned and etched by RIE ($SF_6$/Ar), (c) after optical lithography steps, an electron beam evaporator was used to deposit Ti(10nm)/Au(100nm) metal contacts and then (d) Ti(20nm)/Pt(20nm)/Au(200nm) metal pads. This last step also allowed the definition of the coplanar waveguides (CPWs) for the RF devices. (e) and (f) show optical images of the RF device integrating CPW and of the DC device, respectively.

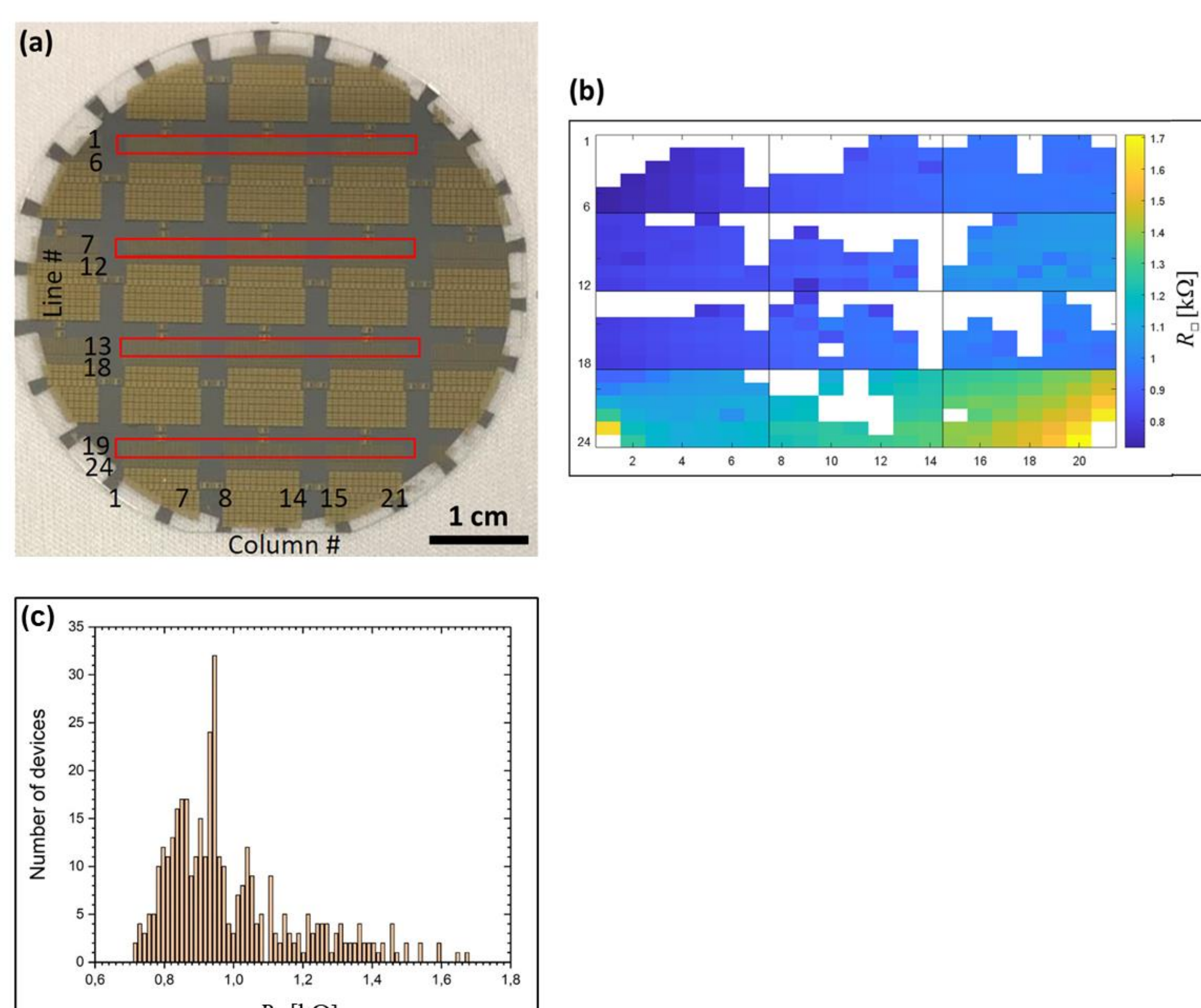

**Figure S13. Statistical measurement of the sheet resistance of DC devices.** (a) $PtSe_2$-based devices fabricated on a 2-inch sapphire substrate with the measured DC devices enclosed in white rectangles. (b) Electrical mapping of the $R_\square$ values showing a clear increase in $R_\square$ for the devices located in lines 19 to 24. These devices make the main contribution to the standard deviation value of 0.20 $k\Omega.\square^{-1}$. (c) Histogram representing the dispersion of the $R_\square$ values over the 378 devices.

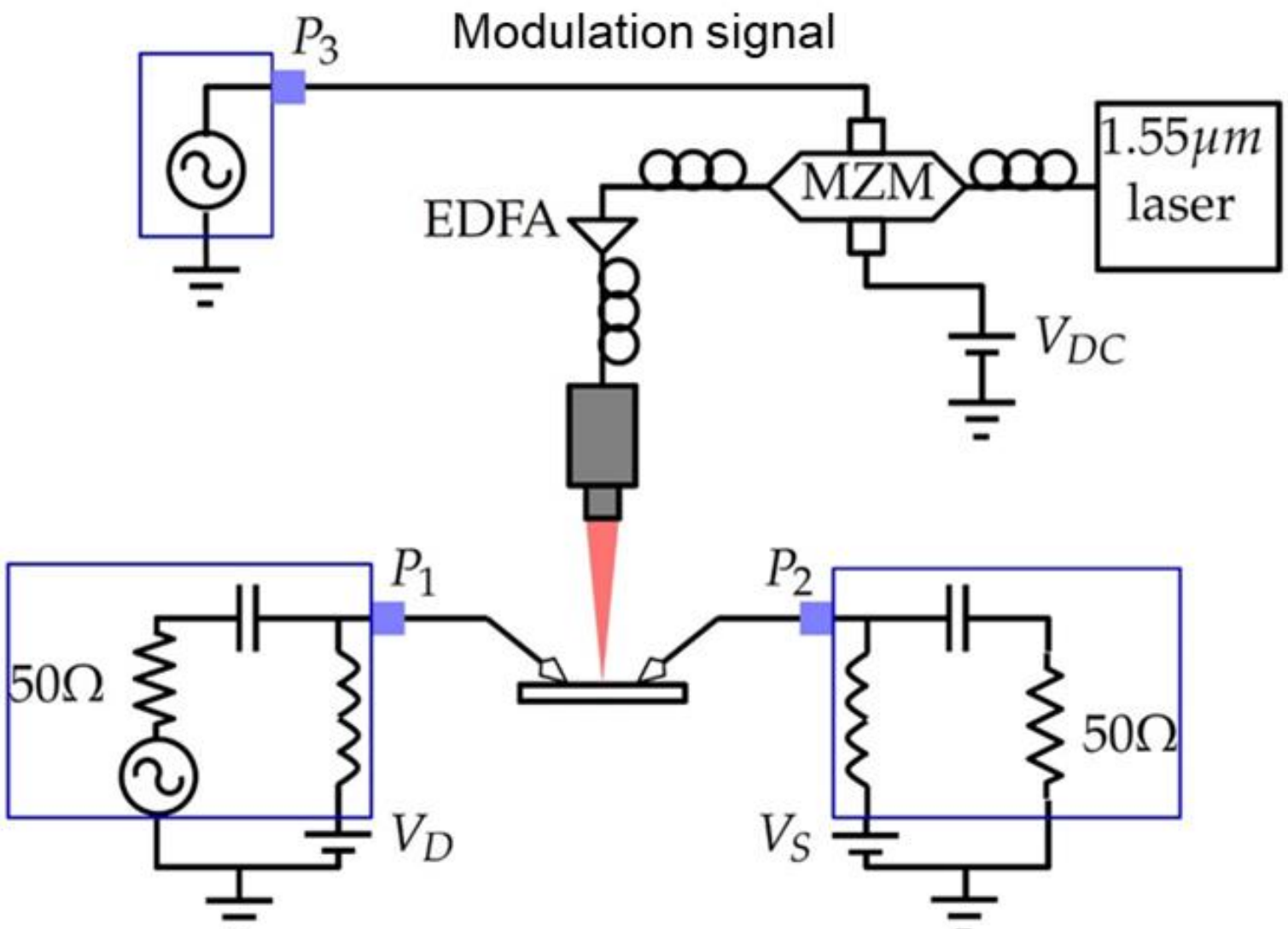

**Figure S14. Detailed schematic of the experimental setup used for photodetection and optoelectronic mixing measurements.** The optical signal is generated by a 1.55 µm laser. Its intensity is modulated by a Mach Zehnder modulator (MZM) using a local oscillator (LO) signal at $f_{opt}$ generated from port 3 (P3) of the virtual network analyzer (VNA). The MZM operation is also controlled by a DC bias ($V_{DC}$). The optical signal is amplified by an erbium-doped fiber amplifier (EDFA) and is focused onto the $PtSe_2$ channel of the device (the laser spot diameter is 2.5 µm). The VNA is internally equipped with a bias T (capacitor and inductor in the schematic) to allow the application of drain-source biases ($V_D$, $V_S$) from external DC source measure units (SMUs). The input RF or DC signal is injected from port 1 (P1) and the mixing and photodetection output signal is received at port 2 (P2).

Photodetection: A RF signal ($f_{opt}$ = 2 - 67 GHz) generated by port 3 (P3) of the VNA is applied to the MZM to modulate the optical signal. $PtSe_2$ acts as a photoconductor thus, the $PtSe_2$ channel is biased through a probe from port 1 (P1) with $V_D$ = 4 V and $V_S$ = 0 V and another RF probe measures the signal generated by the device at the same frequency ($f_{opt}$) at port 2 (P2) of the VNA.

Optoelectronic mixing: $f_{opt}$ is fixed at 30 GHz. A RF signal ($f_{RF}$ = 0.01 - 29.99 GHz) generated by port 1 (P1) of the VNA is applied to the device using an RF probe. Another RF probe measures the signal generated by the device in the intermediate frequency band ($f_{IF} = f_{opt} - f_{RF}$) at port 2 (P2) of the VNA. As the $PtSe_2$ channel is biased by the input RF signal, no DC bias is required: $V_S = V_D = 0$ V.

For both RF measurements, a ZVA67 system from Rohde&Schwarz is used. The system has been calibrated up until the RF probes. The light is modulated using a MXAN-LN-40 MZM from iXBlue Photonics. For $f_{opt}$ up to 33.5 GHz, the system is used in quadrature mode (transmission = 50 %), where the light modulation frequency is the RF frequency from P3 ($f_{RF3}$) applied to the MZM: $f_{opt} = f_{RF3}$. For $f_{opt}$ between 33.5 GHz and 67 GHz, we use the minimum mode (transmission = 0 %), where the light modulation results from the beatings between the optical components at $f_0 - f_{RF3}$ and $f_0 + f_{RF3}$, (where $f_0$ is the 1.55 µm light frequency), thus the light is modulated at twice the applied RF frequency: $f_{opt} = 2 \times f_{RF3}$.

| **Material** | **Operating $\lambda$ (µm)** | **Bandwidth (GHz)** | **Conversion efficiency (dB)** | **Reference** |
|---|---|---|---|---|
| **Graphene (transistor)** | 1.55 | 20 (RF)<br>> 67 (optical) | - 67 | Montanaro *et al.* 2021 [11] |
| **Graphene** | 1.55 | > 30 | - 85 | Montanaro *et al.* 2016 [12] |
| **$PtSe_2$: this work** | 1.55 | > 30 | - 73 | This work |
| **LT-GaAs** | 1.55 | > 67 | - 66 | Billet *et al.* 2017 [13] |

**Table S3: Comparison of 1.55 µm OEMs based on 2D materials and low-temperature (LT) GaAs.**